\documentclass{aastex62}

\usepackage{booktabs}
\usepackage{multirow}
\usepackage{graphicx}
\usepackage{gensymb}
\usepackage{txfonts}
\usepackage{longtable}

\usepackage{mathrsfs}         

\received{January 1, 2018}
\revised{January 7, 2018}
\accepted{July 9, 2018}
\submitjournal{ApJ}

\shorttitle{A study of the  $c$-$\mathrm{C_{3}HD}$/$c$-$\mathrm{C_{3}H_{2}}$ ratio in low-mass star forming regions}
\shortauthors{Chantzos et al.}

\begin{document}

\title{A study of the  $c$-$\mathrm{C_{3}HD}$/$c$-$\mathrm{C_{3}H_{2}}$ ratio in low-mass star forming regions \thanks{Based on observations carried out with the IRAM 30m Telescope. IRAM is supported by INSU/CNRS (France), MPG (Germany), and IGN (Spain)}}

\correspondingauthor{J. Chantzos}
\email{chantzos@mpe.mpg.de}

\author{J. Chantzos}
\affil{Center for Astrochemical Studies, 
           Max-Planck-Institut f\"ur extraterrestrische Physik \\
    Gie\ss enbachstra\ss e 1 \\
85748 Garching (Germany)}

\author{S. Spezzano}
\affil{Center for Astrochemical Studies, 
           Max-Planck-Institut f\"ur extraterrestrische Physik \\
    Gie\ss enbachstra\ss e 1 \\
85748 Garching (Germany)}

\author{P. Caselli}
\affil{Center for Astrochemical Studies, 
           Max-Planck-Institut f\"ur extraterrestrische Physik \\
    Gie\ss enbachstra\ss e 1 \\
85748 Garching (Germany)}

\author{A. Chac\'{o}n-Tanarro}
\affil{Center for Astrochemical Studies, 
           Max-Planck-Institut f\"ur extraterrestrische Physik \\
    Gie\ss enbachstra\ss e 1 \\
85748 Garching (Germany)}

\author{L. Bizzocchi}
\affil{Center for Astrochemical Studies, 
           Max-Planck-Institut f\"ur extraterrestrische Physik \\
    Gie\ss enbachstra\ss e 1 \\
85748 Garching (Germany)}

\author{O. Sipil\"a}
\affil{Center for Astrochemical Studies, 
           Max-Planck-Institut f\"ur extraterrestrische Physik \\
    Gie\ss enbachstra\ss e 1 \\
85748 Garching (Germany)}

\author{B. M. Giuliano}
\affil{Center for Astrochemical Studies, 
           Max-Planck-Institut f\"ur extraterrestrische Physik \\
    Gie\ss enbachstra\ss e 1 \\
85748 Garching (Germany)}

\begin{abstract}
We use the deuteration of $c$-$\mathrm{C_{3}H_{2}}$ to probe the physical parameters of starless and protostellar cores, related to their evolutionary states, and compare it to the $\mathrm{N_{2}H^{+}}$-deuteration in order to study possible differences between the deuteration of C- and N-bearing species.
We observed the main species  $c$-$\mathrm{C_{3}H_{2}}$, the singly and doubly deuterated species $c$-$\mathrm{C_{3}HD}$ and $c$-$\mathrm{C_{3}D_{2}}$, as well as the isotopologue $c$-$\mathrm{{H^{13}CC_{2}H}}$  toward 10 starless cores and 5 protostars in the Taurus and Perseus Complexes. We examined the correlation between the \linebreak $N$($c$-$\mathrm{C_{3}HD}$)/$N$($c$-$\mathrm{C_{3}H_{2}}$) ratio and the dust temperature along with the $\mathrm{H_2}$ column density and the CO depletion factor. The resulting $N$($c$-$\mathrm{C_{3}HD}$)/$N$($c$-$\mathrm{C_{3}H_{2}}$) ratio is within the error bars consistent with 10\% in all starless cores with detected $c$-$\mathrm{C_{3}HD}$. This also accounts for the protostars except for the source HH211, where we measure a high deuteration level of $23\%$. The deuteration of $\mathrm{N_{2}H^{+}}$ follows the same trend but is considerably higher in the dynamically evolved core L1544. We find no significant correlation between the deuteration of $c$-$\mathrm{C_{3}H_{2}}$ and the CO depletion factor among the starless and the protostellar cores. Toward the protostellar cores the coolest objects show the largest deuterium fraction in $c$-$\mathrm{C_{3}H_{2}}$. 
We show that the deuteration of $c$-$\mathrm{C_{3}H_{2}}$ can trace the early phases of star formation and is comparable to that of $\mathrm{N_{2}H^{+}}$.
However, the largest $c$-$\mathrm{C_{3}H_{2}}$ deuteration level  is found toward protostellar cores, suggesting that while $c$-$\mathrm{C_{3}H_{2}}$ is mainly frozen onto dust grains in the central regions of starless cores, active deuteration is taking place on ice.

\end{abstract}


\section{Introduction} \label{intro}
During the early stages of star formation, self-gravitating starless cores begin to contract  ($n_{\mathrm{H}} > 10^4 \, \mathrm{cm^{-3}}$) and to cool down to a few Kelvin ($T<10$ K). 
Under these conditions, the deuteration of molecules is considerably increasing.  $\mathrm{H_{2}D^{+}}$ is the main deuterium provider for most molecules in dense cores and is formed from the following deuteron-proton reaction \citep{millar}:
\begin{eqnarray} \label{deuteration}
\mathrm{H_{3}^{+} + HD \rightleftharpoons H_{2}D^{+} + H_{2} + 230 \, K.}
\end{eqnarray}
 
This reaction is exothermic and proceeds mostly in the forward direction at  temperatures lower than 30 K, increasing the abundance of  
$\mathrm{H_{2}D^{+}}$. The abundance of $\mathrm{H_{2}D^{+}}$ also depends on the ortho-to-para-$\mathrm{H_{2}}$ ratio. The backward direction of reaction \ref{deuteration} is endothermic if the reactants $\mathrm{H_{2}D^{+}}$ and $\mathrm{H_{2}}$ are mostly in the para form \citep{pagani}. An important process in these environments is the freeze-out of molecules onto dust grains. Previous studies have shown that CO depletes heavily toward the center of cores \citep{willacy,caselli99, bacmann, crapsi1, pagani07}. Since CO destroys $\mathrm{H_3^{+}}$ and $\mathrm{H_{2}D^{+}}$, its depletion from the gas phase leads to a further enhancement of the total deuteration level \citep{dalgarno}. 
As the starless core continues to contract, it eventually becomes a pre-stellar core, defined as being a self-gravitating core \citep{ward} with signs of contraction motions and high levels of CO freeze-out and deuteration \citep{crapsi1}. Pre-stellar cores are a sub-sample of starless cores, i.e. the most dynamically evolved and destined to form one or more protostars. Once the protostar is formed, the central regions of the contracting core warm up and the whole core starts to become affected by the outflow driven by the young stellar object. This causes CO to desorb and also increases the backward rate of reaction \ref{deuteration}, leading finally to a decrease of the total deuteration degree. 
In summary,  the increase and decrease of molecular deuteration is sensitive to the evolutionary stage in the star formation process and is an excellent tool to trace the early stages of star formation. 

Species that can be used as evolutionary tracers of high-density and low-temperature gas must have the possibility of deuteration and should be abundant in space. One molecule that has been proven to be a very good evolutionary tracer is diazenylium, $\mathrm{N_{2}H^{+}}$ \citep{crapsi1, emprecht, friesen, punanova}.  It has been shown that the deuteration of  $\mathrm{N_{2}H^{+}}$  correlates tightly with important evolutionary indicators, such as the dust temperature, the CO depletion factor and the central column density of $\mathrm{H_{2}}$. In addition, the emission maps of $\mathrm{N_{2}H^{+}}$  strongly follow the dust emission maps, indicating that $\mathrm{N_{2}H^{+}}$ is less depleted than C-bearing molecules at higher densities \citep{bergin, tafalla}. This makes $\mathrm{N_{2}H^{+}}$ a very good tracer of the deuteration level in central regions of dense cores. The resistance of $\mathrm{N_{2}H^{+}}$ to depletion has been ascribed by previous studies \citep{flower, legal} to the fact that nitrogen in the ISM is mainly in atomic form, and N atoms could stay in the gas phase longer because of their lower sticking probabilities and the slow process which transforms N into $\mathrm{N_{2}}$. 

Following the detection of $c$-$\mathrm{C_{3}H_{2}}$ in the laboratory \citep{thaddeus1}, a number of U-lines observed by \cite{thaddeus2} could be identified as $c$-$\mathrm{C_{3}H_{2}}$ transitions.  After its first detection,  cyclopropenylidene has been observed in various sources, like cold dark clouds, diffuse clouds, circumstellar envelopes, planetary nebulae etc. \citep[e.g.][and references therein]{benson, spezzano1}. Due to the high abundance of the normal species, also the singly deuterated species $c$-$\mathrm{C_{3}HD}$ and the isotopologue $c$-$\mathrm{{H^{13}CC_{2}H}}$ have been observed in various sources in the past 30 years. For example,  \mbox{$c$-$\mathrm{C_{3}HD}$} was detected toward L1498 \citep{bell}, \mbox{TMC-1C} and L1544 \citep{spezzano1} as well as TMC-1 \citep{turner} in the Taurus Molecular Cloud. 

After the first laboratory measurement of the doubly deuterated species by \cite{spezzano2},  $c$-$\mathrm{C_{3}D_2}$ was detected in the ISM  for the first time toward the starless cores TMC-1C and L1544 with a high signal-to-noise-ratio \citep{spezzano1}. The observed single and double deuteration in these two sources could be reproduced by a chemical model including only gas-phase reactions \citep{aikawa}. The possibility of double deuteration and its gas-phase chemistry makes $c$-$\mathrm{C_{3}H_2}$ a useful probe for the deuteration processes taking place only in the gas phase.  A study of the $c$-$\mathrm{C_{3}H_2}$-deuteration as an evolutionary indicator will give complementary information to the deuteration of $\mathrm{N_{2}H^{+}}$, and reveal possible differences between the deuteration of C- and N-bearing species in the gas phase. In fact, unlike $\mathrm{N_{2}H^{+}}$, $c$-$\mathrm{C_{3}H_2}$ is believed to be an early-type molecule \citep{herbst} in terms of cloud evolution and can therefore trace the early stages of star formation;  in particular, in L1544, $c$-$\mathrm{C_{3}H_2}$ preferentially traces the side of the core more exposed to the interstellar radiation field, where the chemistry is continually rejuvenated by the photodissociation of CO \citep{spezzano16}. However, $c$-$\mathrm{C_{3}H_2}$ is also affected by freeze-out or chemical depletion, as it does not trace the central regions of starless cores  
\citep[e.g.] [] {spezzano17}. It is therefore important to compare the deuteration of this molecule in starless and protostellar cores, in order to investigate the level of deuteration in different stages of low-mass star formation. This can give us insights on possible deuteration processes taking place on grain surfaces during the cold and dense phases just before the switch-on of the protostar.

In this work we present single pointing observations of $c$-$\mathrm{C_{3}H_{2}}$, the singly and doubly deuterated species, $c$-$\mathrm{C_{3}HD}$ and $c$-$\mathrm{C_{3}D_{2}}$, as well as the isotopologue $c$-$\mathrm{{H^{13}CC_{2}H}}$ toward 10 starless cores and 5 protostars in the Taurus and Perseus Molecular Cloud Complexes. 
In Section \ref{obs} we summarize the details concerning the observations. Section \ref{results} describes the calculation of the single and double deuteration of $c$-$\mathrm{C_{3}H_{2}}$ as well as the comparison between the deuterium fraction of $\mathrm{N_{2}H^{+}}$ and $c$-$\mathrm{C_{3}H_{2}}$. In Section \ref{results} we also describe the correlation between the deuteration level of $c$-$\mathrm{C_{3}H_{2}}$  and important evolutionary indicators (dust temperature, CO depletion level and central column density of $\mathrm{H_{2}}$). The conclusions are summarized in Section \ref{con}.

\section{Observations} \label{obs}
The observations were done at the IRAM 30m telescope located at Pico Veleta (Spain) toward 10 starless cores and 5 protostellar cores in Taurus and Perseus.  A summary of  the observed objects, their coordinates and their distances is reported in Table \ref{tab:coordinates}. These sources lie all in our Galactic vicinity and represent different stages of star formation, from starless cores, to more evolved pre-stellar cores and young Class 0 protostars.
Previous observations of these sources showed a significant deuteration of  $\mathrm{N_{2}H^{+}}$ which correlates with evolutionary indicators, such as the dust temperature, the CO depletion and central column density of $\mathrm{H_2}$ \citep{crapsi1, emprecht}. 
A special case in our sample of sources is L1521F. Even though some studies \citep{crapsi1} describe this source as an evolved pre-stellar core, successive work  \citep{bourke, Takahashi} has proven that L1521F shows an infrared source and a compact continuum millimeter emission which indicates the existence of a protostar. The detection of a small outflow and a low bolometric luminosity (0.034-0.07 $\mathrm{L_{\odot}}$) suggests that this source is a so-called Very Low Luminosity Object (hereafter VeLLO)  which could be a very young protostellar source or a protostar at its minimum of activity, if episodic accretion is at work \citep[e.g.] [] {Visser}.   \\
The observations were carried out with the EMIR receiver using the E090 configuration (3mm atmospheric window). Each sub-band covered a frequency range of 1.8 GHz, leading to a total spectral coverage of 7.2 GHz.  All four EMIR sub-bands were connected to the Fast Fourier Transform Spectrometer with a frequency resolution of 50 kHz. Frequency switching was performed with a frequency throw of $\pm$3.9 MHz. Telescope pointing was checked every 2 hr on Mercury and was accurate to 3$''$-4$''$.  The intensity of the obtained spectra was given in antenna temperature units, $T^{*}_{\mathrm{A}}$. The antenna temperature  $T^{*}_{\mathrm{A}}$ was converted to the main beam temperature $T_{\mathrm{mb}}$ by using the relation $T_{\mathrm{mb}} = \frac{F_{\mathrm{eff}}}{B_{\mathrm{eff}}} \cdot {T^{*}_{\mathrm{A}}}$. 

 In both samples we observed the main isotopologue  $c$-$\mathrm{C_{3}H_{2}}$, the singly and doubly deuterated species $c$-$\mathrm{C_{3}HD}$ and $c$-$\mathrm{C_{3}D_{2}}$  as well as  the isotopologue $c$-$\mathrm{{H^{13}CC_{2}H}}$ with one $^{13}\mathrm{C}$  being off the principal axis of the molecule. Table \ref{tab:spec_parameters} summarizes the observed species,  the  spectroscopic parameters and telescope settings at the corresponding frequencies: $E_{\mathrm{up}}$ describes the upper state energy and $A_{ul}$ is the Einstein coefficient of the corresponding transition. The upper state degeneracy is given by $g_u$. The parameters $B_{\mathrm{eff}}$ and $\theta_{\mathrm{MB}}$ describe the main beam efficiency and the main beam size of the telescope at a given frequency, respectively. The system temperature $T_{\mathrm{sys}}$  is given in K. The forward efficiency, $F_{\mathrm{eff}}$, is in the observing frequency range equal to 95\%.

\begin{deluxetable*} {cccccc}
\tablewidth{0.0pt}
\tablecaption{Observed sources with their corresponding coordinates and estimated distances. Here we also include the well studied objects L1544 and IRAS16293-2422. \label{tab:coordinates}} 
\tablecolumns{6}
\tablehead{\colhead{Source} & \colhead{Object} & \colhead{R.A} & \colhead{Decl.}  &  \colhead{Distance} & \colhead{References} \\
& & \colhead{J2000.0} & \colhead{J2000.0} & (pc) &}
\startdata
L1495 & starless core  & 04 14 08.2  & +28 08 16 & 140 &  2 \\ 
L1495B &  starless core & 04 18 05.1 & +28 22 22 & 140 & 1 \\
L1495AN &  starless core & 04 18 31.8 & +28 27 30 & 140 & 1\\
L1495AS &  starless core & 04 18 41.8 & +28 23 50 &140 & 1\\
L1400K &  starless core & 04 30 52.1 & +54 51 55 & 140 & 2\\
L1400A &  starless core & 04 30 56.8 & +54 52 36 &140 & 1\\
CB23 &  starless core  & 04 43 27.7 & +29 39 11 & 140 & 1 \\
L1517B &  starless core & 04 55 18.8 & +30 38 04 & 140 & 1\\
L1512 & starless core & 05 04 09.7 & +32 43 09 & 140 & 1\\
TMC2 &  pre-stellar core\tablenotemark{a}  & 04 32 48.7 & +24 25 12 & 140 & 1 \\
L1544 &   pre-stellar core\tablenotemark{a} & 05 04 17.2 & +25 10 43 & 140 & 3 \\
L1521F & protostellar core & 04 28 39.8 & +26 51 35 & 140 & 1\\
Per 5 &  protostellar core & 03 29 51.6 & +31 39 03 & 220 & 4 \\
IRAS03282 &  protostellar core & 03 31 21.0 & +30 45 30 &220 & 4\\
HH211 &  protostellar core & 03 43 56.8 & +32 00 50 & 220 & 4 \\
L1448IRS2 &  protostellar core & 03 25 22.4  & +30 45 12 & 220 & 4 \\
IRAS16293 &  protostellar core & 16 32 22.6  & -24 28 33 & 120 & 5 \\
\enddata
\tablerefs{(1) \cite{lee}; (2) \cite{tafalla}; (3) \cite{ward}; (4) \cite{cernis}; (5) \cite{caux}}
\tablenotetext{a}{The definition of a "pre-stellar core" is based on the work of \cite{crapsi1}.}
\end{deluxetable*}

\section{Results} \label{results}
In the Appendix A (Figures \ref{Fig:ratios1}- \ref{Fig:ratios16}) we show the observed spectra toward all sources. 
The  $3_{2,2}-3_{1,3}$ transition of $c$-$\mathrm{C_{3}H_{2}}$ at 84.727 GHz was detected in all starless and pre-stellar cores except for L1400K, L1517B and L1512. The same line of $c$-$\mathrm{C_{3}H_{2}}$ was also detected in all protostars except for L1448IRS2. The $c$-$\mathrm{C_{3}HD}$ ($3_{0,3}-2_{1,2}$)  and  $c$-$\mathrm{{H^{13}CC_{2}H}}$ ($2_{1,2}-1_{0,1}$) emission was detected in all starless, pre-stellar and protostellar cores with a very high S/N; the strongest lines have a S/N ratio of 25 and 40, respectively. The $3_{1,3}-2_{0,2}$ transition of  $c$-$\mathrm{C_{3}D_{2}}$ at 97.761 GHz
was found at a high S/N level (the strongest line was detected at a 27$\sigma$ level) in the following starless and pre-stellar cores: L1495AN, L1512, L1517B, TMC2 and L1544. In case of the protostellar cores, \mbox{$c$-$\mathrm{C_{3}D_{2}}$} was detected in four sources:  Per5, HH211, L1521F and IRAS16293-2422 with S/N levels ranging from 7.8 to 13.5.

\begin{deluxetable*} {c c c c c c c c c c} 
\tablewidth{0.0pt}
\tablecaption{Transition parameters of the observed species and telescope settings. \label{tab:spec_parameters}}
\tablecolumns{10}
\tablehead{\colhead{Species} & \colhead{Transitions} & \colhead{Frequency} &  \colhead{Ref.} &  \colhead{$E_{\mathrm{up}}$}  & \colhead{$A_{ul}$} & \colhead{$g_u$} & \colhead{$B_{\mathrm{eff}}$} & \colhead{$\theta_{\mathrm{MB}}$} & \colhead{$T_{\mathrm{sys}}$} \\
& $\mathrm{J}_{K_{a},K_{c}}$ & (GHz) & &  ($ \mathrm{K}$) & ($10^{-5} \mathrm{s^{-1}}$)  & & (\%) & ($''$) & (K)}
\startdata
\hline \\ [-1ex]
$c$-$\mathrm{C_{3}H_{2}}$ & $3_{2,2}-3_{1,3}$  & 84.727 & 1& 16.14 & 1.04  & 7 & 81 & 29 & 80-120\\
$c$-$\mathrm{C_{3}HD}$ & $3_{0,3}-2_{1,2}$  & 104.187 & 2 & 10.85 & 3.95 & 21 & 79 & 25 & 90-151 \\
$c$-$\mathrm{C_{3}HD}$ & $2_{1,1}-1_{1,0}$  & 95.994 & 2 & 7.56 & 0.45 & 15 & 80 & 27 & 99-114 \\
$c$-$\mathrm{{H^{13}CC_{2}H}}$ & $2_{1,2}-1_{0,1}$  & 84.185 & 2 & 6.33 & 2.17 & 10 & 81 & 29 & 80-116\\
$c$-$\mathrm{C_{3}D_{2}}$ & $3_{0,3}-2_{1,2}$  & 94.371 & 3 &  9.85 & 3.37 & 21 & 80 & 27 & 82-123\\
$c$-$\mathrm{C_{3}D_{2}}$ & $3_{1,3}-2_{0,2}$  & 97.761 & 3 & 9.88 & 3.87  & 42 & 80 & 26 & 79-119\\
$c$-$\mathrm{C_{3}D_{2}}$ & $2_{2,1}-1_{1,0}$  & 108.654 & 3 &  7.90 & 4.79 & 15 & 78 & 24 & 96-105 \\
\hline
\enddata
\tablerefs{(1) \cite{thaddeus1}; (2) \cite{bogey}; (3) \cite{spezzano2}}
\end{deluxetable*}

\subsection{Calculation of the column densities and the deuteration level} \label{calculation}
The data reduction and analysis was done using the GILDAS\footnote{http://www.iram.fr/IRAMFR/GILDAS} software \citep{pety}. In order to substract the baseline caused mainly by the frequency switching mode, high-order polynomials were fitted. Each line was fitted by using the standard CLASS gaussian fitting method. The total column density was calculated by using the expression for optically thin transitions:
\begin{eqnarray} \label{Eq:column_density_thin}
N_{\mathrm{tot}} = \frac{8 \pi k_{B} W \nu^2 }{A_{ul} h c^3} \cdot \frac{J({T_{\mathrm{ex}}})}{J({T_{\mathrm{ex}}})-J({T_{\mathrm{bg}}})} \cdot  \frac{Q_{\mathrm{rot}}(T_{\mathrm{ex}})}{{g_u} e^{-\frac{E_{\mathrm{up}}}{k_{B} T_{\mathrm{ex}}}}},  
\end{eqnarray}
\noindent{where $W=\frac{\sqrt{\pi} \Delta \varv T_{\mathrm{mb}}}{2 \sqrt{\ln(2)}}$ is the integrated intensity of the line, with $\Delta \varv$ being the linewidth (FWHM)} \citep{caselli2002}. $J(T) = (\frac{h \nu}{k_B})(e^{\frac{\mathrm{h \nu}}{k_{B} T }}-1)^{-1}$ is the Rayleigh-Jeans temperature in K,  $k_B$ is the Boltzmann constant, $\nu$ is the transition frequency, $c$ is the speed of light, and $h$ is the Planck constant. The partition function of a molecule at a given excitation temperature $T_{\mathrm{ex}}$ is given by $Q_{\mathrm{rot}}$.  $T_{\mathrm{bg}}$ is the cosmic background temperature (2.7 K). For a further extension of our sample we also included in this study the pre-stellar core L1544 \citep{spezzano1} as well as the Class 0 protostar IRAS16293-2422 (hereafter IRAS16293) \citep{caux}.  In Tables \ref{tab:line_chararcteristics} and \ref{tab:line_chararcteristics_protostars} we summarize the detected lines in every source and the line properties which were derived from Gaussian fits. The sources CB23, L1495AN, L1495B, Per5, HH211, L1448IRS2, IRAS03282 and IRAS16293 show differences in the linewidth between the main isotopologue and the isotopic variants ranging in the 0.1-0.5 $\mathrm{km/s}$ interval. These differences do not exhibit any clear trend and are poorly constrained with errors varying from 36\% to 94\%. Such discrepancies are likely to be produced by the high noise levels of these latter observations and by the coarse sampling of the line profiles (channel spacing is 0.167 $\mathrm{km/s}$).

\newpage

\startlongtable
\begin{deluxetable*}  {c c c c c c c} 
\tablewidth{0.0pt}
\tablecaption{Observed lines in the starless core sample. The line properties are derived from Gaussian fits. \label{tab:line_chararcteristics}}
\tablehead{\colhead{Source/Molecule} & \colhead{Frequency} & \colhead{$T_{\mathrm{mb}}$} & \colhead{rms} & \colhead{$W$} & \colhead{$\varv_{\mathrm{LSR}}$} & \colhead{$\Delta \varv$} \\
& \colhead{(GHz)} & \colhead{(K)} & \colhead{($\mathrm{m K}$)} &  \colhead{($\mathrm{K \, km \, s^{-1}}$)} & \colhead{($\mathrm{km \, s^{-1}}$)} & \colhead{($\mathrm{km \, s^{-1}}$)}}   
\startdata  
\textbf{CB23} &  &   & &   &   & \\
$c$-$\mathrm{C_{3}H_{2}}$ & 84.727 & 0.05 & 11 & $0.017 \pm 0.004$ & $6.018 \pm 0.037$ & $0.333 \pm 0.088$\\
$c$-$\mathrm{{C_{3}HD}}$ & 104.187 & 0.23 & 9 & $0.056 \pm 0.002$ & $6.015 \pm 0.005$ &  $0.229 \pm 0.010$\\
$c$-$\mathrm{{H^{13}CC_{2}H}}$ & 84.185 & 0.13 & 3 &  $0.024 \pm 0.001$ & $5.987 \pm 0.003$ & $ 0.174 \pm 0.017$\\
$c$-$\mathrm{C_{3}D_{2}}$ & 97.761 & $< 0.02$  & 7 & $< 0.004$  & & \\
\hline \\   [-1ex]
\textbf{L1400A} & & & & &\\ 
$c$-$\mathrm{C_{3}H_{2}}$ & 84.727 & 0.07 & 8 & $0.019 \pm 0.002$ & $3.355 \pm 0.020$ & $0.246 \pm 0.030$ \\  
$c$-$\mathrm{{C_{3}HD}}$ & 104.187 & 0.13 &  8 & $0.036 \pm  0.002$ & $3.348 \pm 0.008$ & $0.263 \pm 0.021$\\
$c$-$\mathrm{{H^{13}CC_{2}H}}$ & 84.185 & 0.06 & 2 & $0.020 \pm 0.001$ & $3.287 \pm 0.006$ & $0.334 \pm 0.014$ \\
 \hline \\  [-1ex]
\textbf{L1400K} & & & & &\\
$c$-$\mathrm{{H^{13}CC_{2}H}}$ & 84.185 & 0.06 & 9 & $0.018 \pm 0.002$ & $3.178 \pm 0.023$ & $0.268 \pm 0.048$\\
$c$-$\mathrm{{C_{3}HD}}$ & 104.187 & $< 0.04$ & 15 &$< 0.01$  & & \\
 \hline \\   [-1ex]
\textbf{L1495} & & & & & \\
$c$-$\mathrm{C_{3}H_{2}}$ & 84.727 & 0.06 & 4 & $0.015 \pm 0.001$ & $6.779 \pm 0.012$ & $0.244 \pm 0.035$ \\
$c$-$\mathrm{{C_{3}HD}}$ & 104.187 & 0.13 & 10 & $0.038 \pm 0.002$ & $6.768 \pm 0.010$ & $0.269 \pm 0.025$ \\
$c$-$\mathrm{{H^{13}CC_{2}H}}$  & 84.185 & 0.07 & 14 &
$0.021 \pm 0.005$ & $6.726 \pm 0.036$ & $0.281 \pm 0.088$\\
\hline \\  [-1ex]
\textbf{L1495AN} & & & & & \\
$c$-$\mathrm{C_{3}H_{2}}$ & 84.727 & 0.17 & 14 & $0.062 \pm 0.006$ &  $7.259 \pm 0.014$ & $0.344 \pm 0.039$\\
$c$-$\mathrm{{C_{3}HD}}$ & 104.187 & 0.43 & 22 & $0.138 \pm 0.007$ & $7.275 \pm 0.007$ & $0.304 \pm 0.019$\\
$c$-$\mathrm{{H^{13}CC_{2}H}}$ &  84.185 & 0.22 & 10 & 
$0.075 \pm 0.004$ & $7.261 \pm 0.007$ &  $0.329 \pm  0.019$\\
$c$-$\mathrm{C_{3}D_{2}}$ & 97.761 & 0.04 & 6 & $0.011 \pm 0.001$ & $7.208 \pm 0.020$ & $0.236 \pm 0.044$\\
\hline \\  [-1ex]
\textbf{L1495AS} & & & & & \\
$c$-$\mathrm{C_{3}H_{2}}$ & 84.727 & 0.08 & 8 & $0.028 \pm  0.004 $ & $7.302 \pm 0.019$ & $0.353 \pm 0.046$\\
$c$-$\mathrm{{C_{3}HD}}$ & 104.187 & 0.15 & 9 & $0.046 \pm  0.002$ & $7.281 \pm 0.008$ & $0.280 \pm 0.018$\\
$c$-$\mathrm{{H^{13}CC_{2}H}}$ &  84.185 & 0.09 & 8 & $0.025 \pm 0.002$ & $7.229 \pm 0.015$ &  $0.254 \pm  0.022$\\
\hline \\  [-1ex]
\textbf{L1495B} & & & & & \\
$c$-$\mathrm{C_{3}H_{2}}$ & 84.727 & 0.04 & 8 & $0.020 \pm 0.004$ & $6.655 \pm 0.038$ & $0.460 \pm 0.086$\\
$c$-$\mathrm{{C_{3}HD}}$ & 104.187 & 0.13 & 16 &  $0.054 \pm 0.006$ & $6.635 \pm 0.022$ & $0.403 \pm  0.048$\\
$c$-$\mathrm{{H^{13}CC_{2}H}}$ & 84.185 & 0.06 & 7 & $0.018 \pm 0.002$  & $6.582 \pm 0.020$ &  $0.300 \pm  0.045$\\
\hline \\  [-1ex]
\textbf{L1512} & & & & & \\
$c$-$\mathrm{C_{3}HD}$ & 104.187 & 0.30 &  19 & $0.067 \pm  0.005$ &  $7.095 \pm 0.009$ & $0.204 \pm 0.015$\\
$c$-$\mathrm{{H^{13}CC_{2}H}}$ & 84.185 & 0.15 & 8 & $0.033 \pm 0.002$ & $7.058 \pm 0.012$ & $0.211 \pm  0.015$\\
$c$-$\mathrm{C_{3}D_{2}}$ & 97.761 & 0.03 & 3 & $0.008 \pm 0.001$ & $7.075 \pm 0.017$ & $0.293 \pm  0.038$\\
 \hline \\   [-1ex]
\textbf{L1517B} & & & & &\\
$c$-$\mathrm{C_{3}HD}$ & 104.187 & 0.30 & 14 & $0.068 \pm  0.004$ & $5.778 \pm  0.007$ &  $0.249 \pm  0.017$\\
$c$-$\mathrm{{H^{13}CC_{2}H}}$ & 84.185 & 0.10 & 11 & 
$0.029 \pm 0.004$  & $5.751 \pm  0.018$ & $0.279 \pm  0.031$\\
$c$-$\mathrm{C_{3}D_{2}}$ & 97.761 & 0.05 & 8 & $0.014 \pm 0.002$ & $5.777 \pm  0.024$ & $0.266 \pm 0.054$\\
$c$-$\mathrm{C_{3}D_{2}}$ & 94.371 & 0.05 & 8 & $0.013 \pm 0.002$ & $5.854 \pm 0.025$  & $0.271 \pm 0.053$\\
 \hline \\  [-1ex]
\textbf{TMC2} & & & & &\\
$c$-$\mathrm{C_{3}H_{2}}$ & 84.727 & 0.16 & 18 &  $0.061 \pm  0.007$ & $6.192 \pm 0.020$  & $0.365 \pm 0.047$ \\
$c$-$\mathrm{C_{3}HD}$ & 104.187 & 0.53 & 24 & $0.184 \pm 0.007$ &  $6.217 \pm 0.007$ & $0.325 \pm 0.016$\\
$c$-$\mathrm{{H^{13}CC_{2}H}}$ & 84.185 & 0.16 & 9 & 
$0.065 \pm  0.004$ & $6.132 \pm 0.010$ & $0.374 \pm  0.024$\\
$c$-$\mathrm{C_{3}D_{2}}$ & 97.761 & 0.10 &  4 & $0.030 \pm  0.001$ &  $6.237 \pm 0.005$  & $0.283 \pm  0.012$\\
$c$-$\mathrm{C_{3}D_{2}}$ & 94.371 & 0.05 & 5 & $0.018 \pm 0.001$ & $6.258 \pm 0.014$ & $0.330 \pm 0.040$\\
$c$-$\mathrm{C_{3}D_{2}}$ & 108.654 & 0.03 & 7 & $0.012 \pm 0.002$ &  $6.262 \pm  0.036$ & $0.355 \pm 0.107$\\
\hline \\   [-1ex]
\textbf{L1544}\tablenotemark{a} & & & & &\\
$c$-$\mathrm{C_{3}H_{2}}$ & 84.727 & 0.21 & 10 &  $0.10 \pm 0.01$ & $7.210 \pm  0.008$ &  $0.46 \pm 0.01$\\
$c$-$\mathrm{C_{3}HD}$ & 104.187 & 0.48 & 10 &  $0.238 \pm 0.004$ & $7.181 \pm  0.004$  &  $0.468 \pm 0.009$\\
$c$-$\mathrm{C_{3}HD}$ & 95.994 & 0.13 & 10 & $0.065 \pm 0.003$ & $7.17 \pm  0.01$  &  $0.48 \pm 0.03$\\
$c$-$\mathrm{{H^{13}CC_{2}H}}$ & 84.185 & 0.19 &  10  & $0.093 \pm 0.003$ & $7.154 \pm  0.008$ & $0.44 \pm  0.02$\\
$c$-$\mathrm{C_{3}D_{2}}$ & 97.761 & 0.13 & 5 & $0.059 \pm 0.002$ & $7.181 \pm  0.007$ & $0.43 \pm 0.02$\\
$c$-$\mathrm{C_{3}D_{2}}$ & 94.371 & 0.07 & 5 &  $0.032 \pm 0.002$ & $7.20 \pm  0.01$ & $0.45 \pm 0.03$\\
$c$-$\mathrm{C_{3}D_{2}}$ & 108.654 & 0.04 & 5 & $0.023 \pm 0.002$ & $7.17 \pm  0.02$ & $0.54 \pm 0.05$
\enddata
\tablenotetext{a}{The values for L1544 were taken from \cite{spezzano1}}. 
\end{deluxetable*}

\begin{deluxetable*} {c c c c c c c} 
\tablewidth{0.0pt}
\tablecaption{Observed lines in the protostellar core sample. The line properties are derived from Gaussian fits. \label{tab:line_chararcteristics_protostars}}
\tablecolumns{10}
\tablehead{\colhead{Source/Molecule} & \colhead{Frequency} & \colhead{$T_{\mathrm{mb}}$} & \colhead{rms} & \colhead{$W$} & \colhead{$\varv_{\mathrm{LSR}}$} & \colhead{$\Delta \varv$}  \\
 & \colhead{(GHz)} & \colhead{(K)} & \colhead{($\mathrm{m K}$)}  & \colhead{($\mathrm{K \, km \, s^{-1}}$)} & \colhead{($\mathrm{km \, s^{-1}}$)} & \colhead{($\mathrm{km \, s^{-1}}$)}}
\startdata
\textbf{Per5} & & & & &\\
$c$-$\mathrm{C_{3}H_{2}}$ & 84.727 & 0.09 & 9 & $0.050 \pm  0.004$ & $8.102 \pm  0.021$ & $0.529 \pm  0.051$\\
$c$-$\mathrm{C_{3}HD}$ & 104.187 & 0.27 & 21 & $0.116 \pm  0.007$ & $8.174 \pm  0.013$ &  $0.401 \pm 0.032$\\
$c$-$\mathrm{{H^{13}CC_{2}H}}$ & 84.185 & 0.09 & 6 & 
$0.035 \pm 0.002$ & $8.171 \pm  0.012$  & $0.362 \pm  0.026$\\
$c$-$\mathrm{C_{3}D_{2}}$ & 97.761 & 0.09 & 11 & $0.029 \pm 0.004$ & $8.215 \pm  0.021$ &  $0.318 \pm 0.050$\\
\hline \\ [-1ex]
\textbf{HH211} & & & & &\\
$c$-$\mathrm{C_{3}H_{2}}$ & 84.727 & 0.10 & 7 & $0.051 \pm 0.002$ & $9.089 \pm 0.013$ &  $0.461 \pm  0.029$\\
$c$-$\mathrm{C_{3}HD}$ & 104.187 & 0.35 & 14 & $0.163 \pm 0.005$ & $9.097 \pm  0.007$  &  $0.432 \pm  0.018$ \\
$c$-$\mathrm{{H^{13}CC_{2}H}}$ & 84.185 & 0.05 & 5 & 
$0.022 \pm  0.002$ & $9.100 \pm 0.019$ & $0.403 \pm  0.045$\\
$c$-$\mathrm{C_{3}D_{2}}$ & 97.761 & 0.10 & 7 & $ 0.037 \pm  0.002$ & $9.102 \pm  0.012$ & $ 0.361 \pm  0.032$\\
\hline \\ [-1ex]
\textbf{L1448IRS2} & & & & &\\
$c$-$\mathrm{C_{3}HD}$ & 104.187 & 0.20 & 13 & $0.105 \pm  0.005$ &  $4.084 \pm  0.011$ & $0.479 \pm  0.027$\\
$c$-$\mathrm{{H^{13}CC_{2}H}}$ & 84.185 & 0.05 & 6 & 
$0.037 \pm  0.004$ & $3.977 \pm  0.032$ & $0.690 \pm  0.082$\\
\hline \\ [-1ex]
\textbf{IRAS03282} & & & & &\\
$c$-$\mathrm{C_{3}H_{2}}$ & 84.727  & 0.02 & 5 &  $0.020 \pm  0.002$ & $7.002 \pm 0.062$ & $0.933 \pm  0.155$\\
$c$-$\mathrm{C_{3}HD}$ & 104.187 & 0.12 & 9 & $0.064 \pm  0.004$ & $6.863 \pm  0.015$ & $0.517 \pm  0.034$\\
$c$-$\mathrm{{H^{13}CC_{2}H}}$ & 84.185 & 0.03 & 5 &  $0.013 \pm 0.002$ & $6.812 \pm  0.035$ &  $ 0.414 \pm  0.072$\\
\hline \\ [-1ex]
\textbf{L1521F} & & & & &\\
$c$-$\mathrm{C_{3}H_{2}}$ & 84.727 & 0.22 & 6 &  $0.086 \pm 0.002$ & $6.407 \pm  0.005$ &  $0.366 \pm 0.010$\\
$c$-$\mathrm{C_{3}HD}$ & 104.187 & 0.33 & 17 & $0.144 \pm 0.006$ & $6.433 \pm  0.009$  &  $0.410 \pm 0.018$\\
$c$-$\mathrm{{H^{13}CC_{2}H}}$ & 84.185 & 0.18 & 12 & $0.077 \pm 0.005$ & $6.365 \pm  0.013$ & $0.404 \pm  0.028$\\
$c$-$\mathrm{C_{3}D_{2}}$ & 97.761 & 0.05 & 6 & $0.018 \pm 0.002$ & $6.477 \pm  0.020$ & $0.360 \pm 0.063$\\
\hline \\ [-1ex]
\textbf{IRAS16293} & & & & &\\
$c$-$\mathrm{C_{3}H_{2}}$ & 84.727  & 0.09 & 3 & $0.183  \pm 0.006$ & $4.303 \pm 0.034$ & $2.011  \pm 0.077$ \\
$c$-$\mathrm{C_{3}HD}$ & 104.187 & 0.17 &  6 & $0.280  \pm 0.012$ & $4.235 \pm  0.031$ & $1.573 \pm  0.080$ \\
$c$-$\mathrm{C_{3}HD}$ & 95.994 & 0.03 & 2 &  $0.053 \pm 0.004$ & $4.026 \pm  0.059$ & $1.914 \pm 0.147$\\
$c$-$\mathrm{{H^{13}CC_{2}H}}$ & 84.185 & 0.03 & 3 & $0.089 \pm  0.008$ & $3.889  \pm  0.108$ &  $2.394 \pm  0.237$\\
$c$-$\mathrm{C_{3}D_{2}}$ & 94.371 & 0.03 & 4 & $0.059 \pm  0.010$ & $4.342  \pm 0.158$ & $1.952 \pm  0.375$ \\
$c$-$\mathrm{C_{3}D_{2}}$ & 97.761 & 0.03 & 3 & $0.049 \pm 0.006$ & $4.281 \pm  0.097$ & $1.516 \pm 0.196$ \\
\hline
\enddata
\end{deluxetable*}

\begin{figure}[h]
	\centering
	\includegraphics[width = 0.5\textwidth]{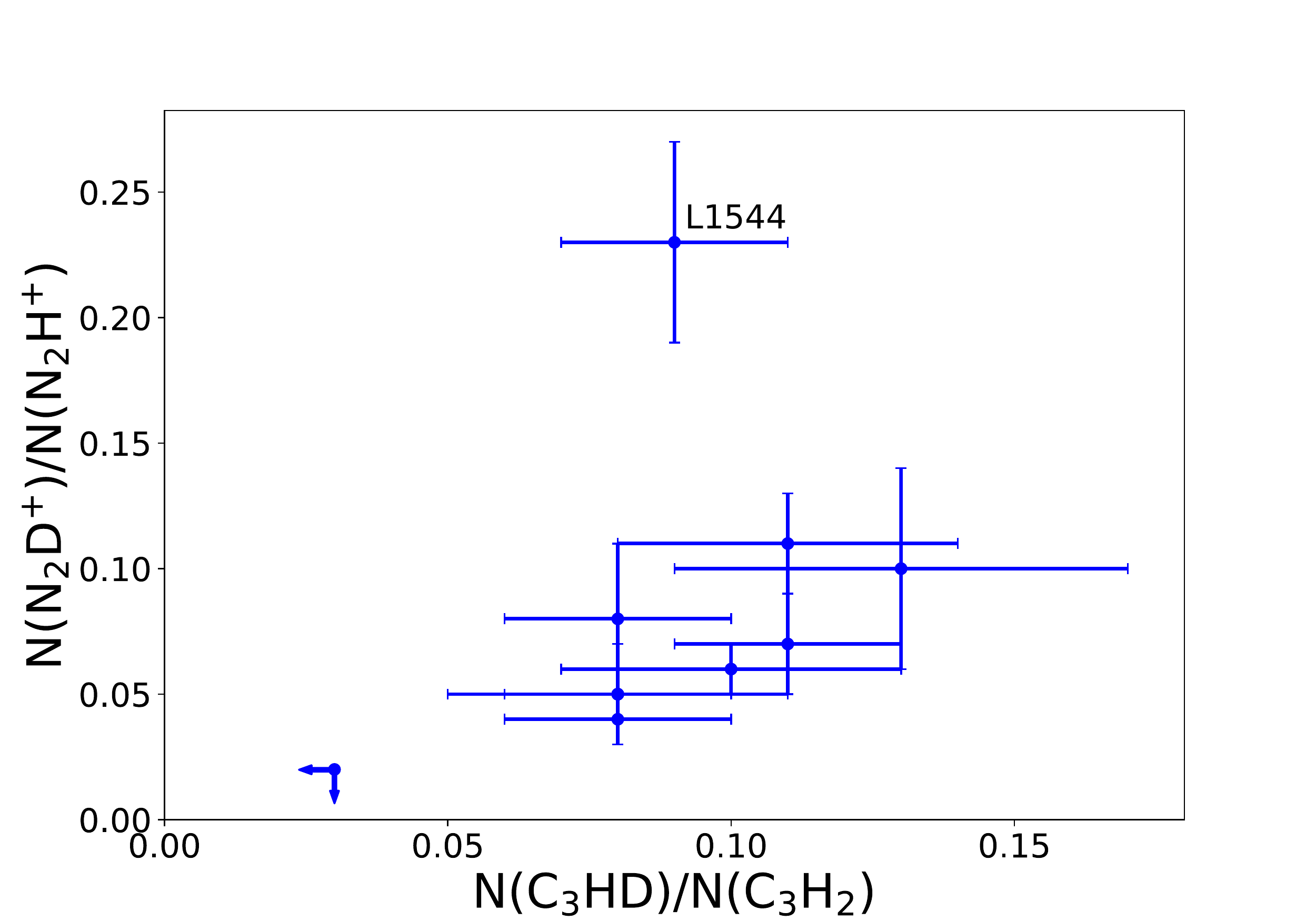}
	\caption{Deuterium fraction of $c$-$\mathrm{C_{3}H_{2}}$ and $\mathrm{N_{2}H^{+}}$ in 11 starless cores located in the Taurus Complex. The deuteration level of $\mathrm{N_{2}H^{+}}$  was calculated in previous work \citep{crapsi1}. The arrows indicate the upper limits for the estimated $c$-$\mathrm{C_{3}H_{2}}$ and $\mathrm{N_{2}H^{+}}$ deuterium fraction in the source L1400K.}
	\label{Fig:pre-stellar_ratio}
\end{figure}

\begin{figure}[h]
	\centering
	\includegraphics[width = 0.5\textwidth]{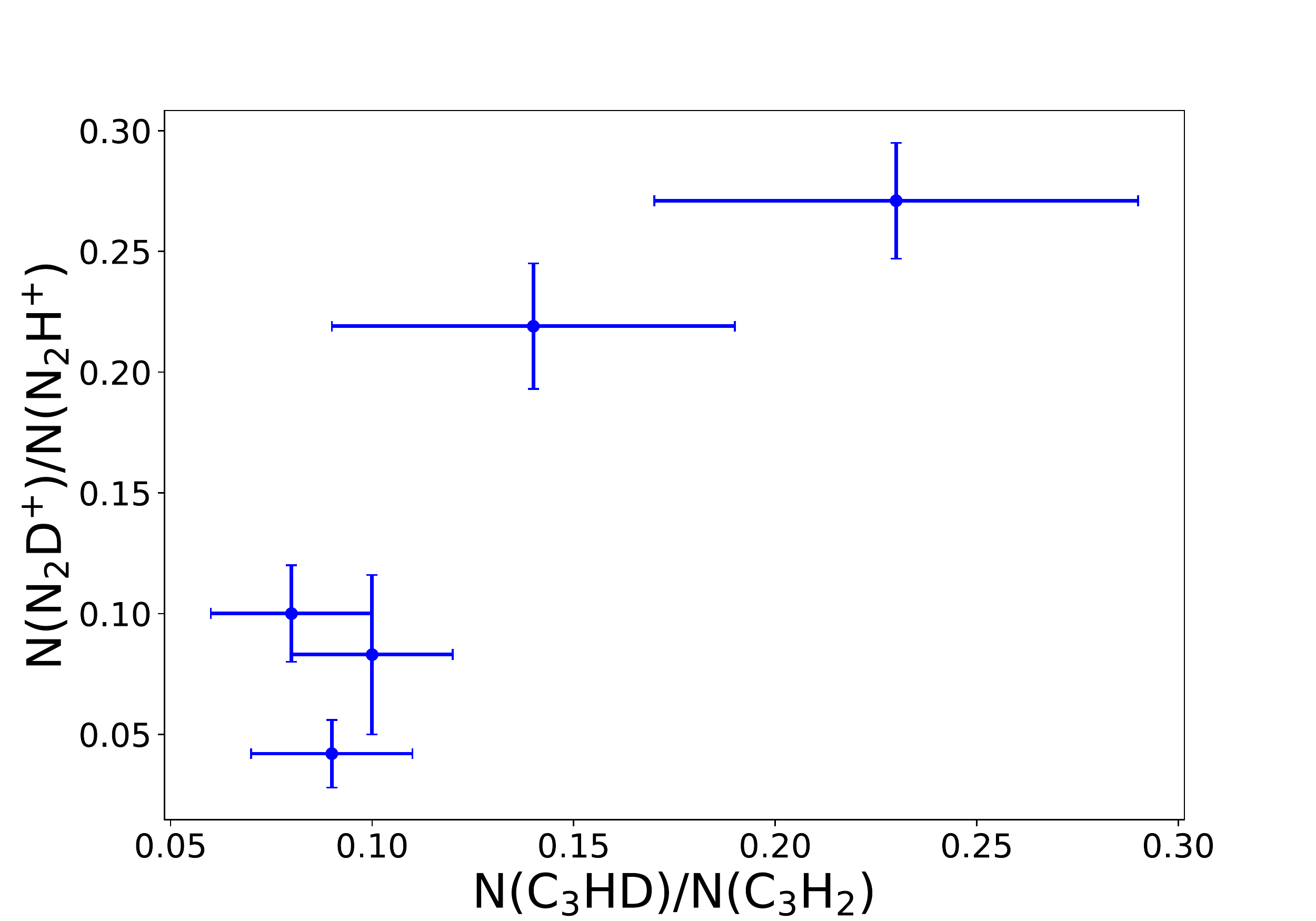}
	\caption{Deuterium fraction of $c$-$\mathrm{C_{3}H_{2}}$ and $\mathrm{N_{2}H^{+}}$ in 4 protostellar cores in the Perseus Complex and one protostellar core (L1521F)  in the Taurus Complex.  The deuteration level of $\mathrm{N_{2}H^{+}}$ was calculated in previous work done by Emprechtinger et al. (2009). }
	\label{Fig:protostar_ratio}
\end{figure}

For the calculation of the total column densities of $c$-$\mathrm{C_{3}H_{2}}$ and its isotopologues we assumed the same excitation temperature  $T_{\mathrm{ex}}$ that has been used in the same sources by \cite{crapsi1} and \cite{emprecht}  for $\mathrm{N_{2}H^{+}}$.\footnote{The critical density of $c$-$\mathrm{C_{3}H_{2}}$ ($3_{2,2}-3_{1,3}$) lies a factor of 12.7 higher than the critical density of $\mathrm{N_{2}H^{+}}$(1-0) at 30 K \citep{chandra, schoier05}.  We cannot compare the critical densities of the above transitions at lower temperatures, since the collisional rate of the $c$-$\mathrm{C_{3}H_{2}}$ ($3_{2,2}-3_{1,3}$) transition is unknown below 30 K.} We also used the same excitation temperature for the deuterated and the $^{13}\mathrm{C}$  species. 
The effect of underestimating  the excitation temperature of the main and the deuterated species by 1 K changes the deuteration level $N$($c$-$\mathrm{C_{3}HD}$)/$N$($c$-$\mathrm{C_{3}H_{2}}$)  up to 30\%.
In case of L1544 we use the $T_{\mathrm{ex}}$ derived in \cite{gerin}, where detections of $c$-$\mathrm{C_{3}H_{2}}$ and its deuterated counterpart are reported. In particular,  there were three transitions of $c$-$\mathrm{C_{3}HD}$ at 19.419, 79.812 and 104.187 GHz detected in L1544 which gave a $T_{\mathrm{ex}}$ of $5 \pm 2$ K.  An optically thick transition of $c$-$\mathrm{C_{3}H_{2}}$ detected at 85.339 GHz provided a $T_{\mathrm{ex}}$ of 6 K. This excitation temperature is, within the errors, equal to that found in $\mathrm{N_{2}H^{+}}$(1-0) and $\mathrm{N_{2}D^{+}}$(2-1) toward the same source \citep{crapsi1}.  Due to the large error of $T_{\mathrm{ex}}$ for $c$-$\mathrm{C_{3}HD}$, we use for the main and the deuterated species a $T_{\mathrm{ex}}$ of 6 K.

Since $c$-$\mathrm{{H^{13}CC_{2}H}}$ has been detected in every source, we used the total column density of $c$-$\mathrm{{H^{13}CC_{2}H}}$ to derive $N$($c$-$\mathrm{C_{3}H_{2}}$) by considering a $^{12}\mathrm{C}$/$^{13}\mathrm{C}$ ratio of 77 \citep{wilson}. This gives us the additional advantage of avoiding ambiguities due to the optical depth of the main species. The carbon isotope ratio has been determined in several sources,  from different molecular species and can vary up to a factor of 2 \citep{wilson}. This means that the derived $N$($c$-$\mathrm{C_{3}H_{2}})$ suffers from an uncertainty of a factor of 2. The assumed ortho to para ratio of $c$-$\mathrm{C_{3}H_{2}}$ and $c$-$\mathrm{C_{3}D_{2}}$ is 3 and 2, respectively. Table \ref{tab:column_densities_pre} and Table \ref{tab:column_densities_pro}  show the derived column densities for every species in each starless core and protostar, respectively. The error for the column densities was calculated by propagating the uncertainty on the integrated intensity, $W$, including the 1$\sigma$ statistical error as well as 10\% calibration uncertainty.
The column densities for the singly and doubly deuterated species were calculated from the $c$-$\mathrm{C_{3}HD}$ ($3_{0,3}-2_{1,2}$) and $c$-$\mathrm{C_{3}D_{2}}$ ($3_{1,3}-2_{0,2}$) transitions. 

Figure \ref{Fig:pre-stellar_ratio}  shows the deuterium fraction for each species in every starless core. Here one can clearly see that the deuteration for both $c$-$\mathrm{C_{3}H_2}$ and $\mathrm{N_{2}H^{+}}$ follows a similar trend and is of the same magnitude, except for the most dynamically evolved object, the pre-stellar core L1544. The deuteration of $c$-$\mathrm{C_{3}H_{2}}$ is in all sources within the 1$\sigma$ uncertainty consistent with 10\%, except for L1400K where the $c$-$\mathrm{C_{3}HD}$ column density was estimated to be less than $0.26 \times 10^{12} \, \mathrm{cm^{-2}}$, which resulted in a 3$\sigma$ upper limit for the $N$($c$-$\mathrm{C_{3}HD}$)/$N$($c$-$\mathrm{C_{3}H_{2}}$) ratio of 0.03.
All these sources except for TMC2 and L1544 have been identified in \cite{crapsi1} as less evolved starless cores, having, among other properties, a deuteration degree $\leq$0.1 which is also in our case fulfilled (in the source L1495B the deuterium fraction can still be within the errors less than or equal to 0.1).  L1544 is the only source where the deuteration of $\mathrm{N_{2}H^{+}}$ is larger than the one of $c$-$\mathrm{C_{3}H_2}$, showing a significant discrepancy of a factor of 2.6. 
One possible explanation for this can be found in the formation route of $\mathrm{N_{2}H^{+}}$. The progenitor of $\mathrm{N_{2}H^{+}}$  is $\mathrm{N_{2}}$ which is formed via neutral-neutral reactions: $\mathrm{N + C \rightarrow CN}$ and $\mathrm{CN + N \rightarrow N_2}$  \citep{pineau, legal}. Carbon-bearing molecules like $c$-$\mathrm{C_{3}H_{2}}$ on the other hand are formed faster through sequential ion-neutral reactions. For this reason,  $\mathrm{N_{2}H^{+}}$ is believed to be a late-type molecule, becoming highly abundant in evolved cores, such as L1544. Furthermore, $\mathrm{N_{2}H^{+}}$ is  more resistant to depletion and survives in the gas phase longer than C-bearing molecules, thus tracing regions where the deuterium fractionation is more efficient because of the large amount of freeze out of neutral species (such as CO and O) which participate in the destruction of the $\mathrm{{H_{3}^{+}}}$ deuterated isotopologues \citep{dalgarno}. 

The abundance of the doubly deuterated species with respect to the normal species is  0.4-1.5\%. These values are comparable to those calculated by \cite{spezzano1} in TMC-1C and L1544. The ratios $N$($c$-$\mathrm{C_{3}HD}$)/$N$($c$-$\mathrm{C_{3}H_{2}}$) and \mbox{$N$($c$-$\mathrm{C_{3}D_{2}}$)/$N$($c$-$\mathrm{C_{3}HD}$)} are quite similar in all starless and pre-stellar cores. This suggests that $c$-$\mathrm{C_{3}H_{2}}$ and $c$-$\mathrm{C_{3}HD}$ follow the same deuteration route and are not affected by the dynamical evolution of the dense core, as already pointed out in \cite{spezzano1}. In the source CB23 we find a marginal detection of $c$-$\mathrm{C_{3}D_{2}}$ and therefore derive a 3$\sigma$ upper limit for the column density. The abundance ratios \mbox{$N$($c$-$\mathrm{C_{3}HD}$)/$N$($c$-$\mathrm{C_{3}H_{2}}$)}, $N$($c$-$\mathrm{C_{3}D_{2}}$)/$N$($c$-$\mathrm{C_{3}H_{2}}$) and \mbox{$N$($c$-$\mathrm{C_{3}D_{2}}$)/$N$($c$-$\mathrm{C_{3}HD}$)} among the starless core sample are listed in Table \ref{tab:ratio_column_densities_pre}.

\begin{deluxetable*}  {c c c c c} 
\tablewidth{0.0pt}
\tablecaption{Column densities of the species $c$-$\mathrm{C_{3}H_{2}}$, $c$-$\mathrm{C_{3}HD}$, $c$-$\mathrm{C_{3}D_{2}}$ and $c$-$\mathrm{{H^{13}CC_{2}H}}$ in the 11 starless cores within the Taurus Molecular Cloud Complex. \label{tab:column_densities_pre}}
\tablehead{\colhead{Starless Core} & \colhead{$N$($c$-$\mathrm{C_{3}H_{2}}$)} & \colhead{$N$ ($c$-$\mathrm{C_{3}HD}$)} & \colhead{$N$($c$-$\mathrm{{H^{13}CC_{2}H}}$)} & \colhead{$N$($c$-$\mathrm{C_{3}D_{2}}$)} \\
& ($10^{12} \, \mathrm{cm^{-2}}$) & ($10^{12} 
\, \mathrm{cm^{-2}}$) & ($10^{12} \, \mathrm{cm^{-2}}$) &  ($10^{12} \, \mathrm{cm^{-2}}$)}
\startdata
CB23 & $10.9 \pm 1.2$ & $1.15 \pm 0.13$ & $0.28 \pm 0.03$ & $<0.07$\\
L1400A & $9.1 \pm 1.1$ & $0.74 \pm 0.09$ & $0.24 \pm 0.03$ & \tablenotemark{-} \\  
L1400K &  $8.0 \pm 1.3 $ & $<0.26$ & $0.21 \pm 0.03$ &  \tablenotemark{-}  \\
L1495 &  $8.6 \pm 2.1$ & $0.66 \pm 0.08$ & $0.22 \pm 0.06$ &   \tablenotemark{-} \\
L1495AN & $31.7 \pm 3.5$ & $2.50 \pm 0.28$ & $0.82 \pm 0.09$ & $0.13 \pm 0.02$ \\
L1495AS & $11.4 \pm 1.6$ & $0.94 \pm 0.11$ & $0.30 \pm 0.04$ &  \tablenotemark{-} \\
L1495B & $7.5 \pm 1.2$ & $0.94 \pm 0.14$ & $0.20 \pm 0.03$ &  \tablenotemark{-} \\
L1512 & $13.2 \pm 1.6$ & $1.09 \pm 0.13$ & $0.34 \pm 0.04$ & $0.08 \pm 0.02$\\
L1517B & $12.2 \pm 1.9$ & $1.23 \pm 0.14$ & $0.32 \pm 0.05$ &  $0.16 \pm 0.03$\\
TMC2 & $25.9 \pm 2.9$ & $2.94 \pm 0.32$ & $0.67 \pm 0.08$ & $0.31 \pm 0.03$\\
L1544 & $34.9 \pm 3.7$ & $3.18 \pm 0.32$ & $0.91 \pm 0.10$ & $0.53 \pm 0.06$\\
\enddata
\tablenotetext{-}{no data available} 
\end{deluxetable*}

\begin{deluxetable*} {c c c c c} 
\tablewidth{0.0pt}
\tablecaption{Column densities of the species $c$-$\mathrm{C_{3}H_{2}}$, $c$-$\mathrm{C_{3}HD}$, $c$-$\mathrm{C_{3}D_{2}}$ and $c$-$\mathrm{{H^{13}CC_{2}H}}$ in the 6 protostars. The protostars Per 5, HH211, L1448IRS2 and IRAS03282 belong to the Perseus Complex, L1521F lies in the Taurus Complex while IRAS16293 is part of the Ophiuchus Complex. \label{tab:column_densities_pro}}
\tablehead{\colhead{Protostellar} & \colhead{$N$($c$-$\mathrm{C_{3}H_{2}}$)} & \colhead{$N$($c$-$\mathrm{C_{3}HD}$)} & \colhead{$N$($c$-$\mathrm{{H^{13}CC_{2}H}}$)} & \colhead{$N$($c$-$\mathrm{C_{3}D_{2}}$)} \\
\colhead{Core} & \colhead{($10^{12} \, \mathrm{cm^{-2}}$)} & \colhead{($10^{12} 
\, \mathrm{cm^{-2}}$)} & \colhead{($10^{12} \, \mathrm{cm^{-2}}$)} &  \colhead{($10^{12} \, \mathrm{cm^{-2}}$)}} 
\startdata
Per5 & $14.2 \pm 1.7$ & $1.4 \pm 0.16$ & $0.37 \pm 0.04$ & $0.24 \pm 0.04$\\
HH211 & $8.5 \pm 1.2$ & $1.9 \pm 0.20$ & $0.22 \pm 0.03$ & $0.30 \pm 0.04$ \\  
L1448IRS2 &  $13.9 \pm 1.9 $ & $1.3 \pm 0.14$ & $0.36 \pm 0.05$ & \tablenotemark{-}  \\
IRAS03282 &  $5.1 \pm 1.1 $ & $0.7 \pm 0.09$ & $0.13 \pm 0.03$ &  \tablenotemark{-} \\
IRAS16293 &  $36.4 \pm 5.0$ & $3.6 \pm 0.39$ & $0.95 \pm 0.13$   & $0.43 \pm 0.07$ \\
L1521F & $31.5 \pm 3.7 $ & $2.4 \pm 0.26$ & $0.82 \pm 0.10$ & $0.20 \pm 0.03$ \\
\enddata
\tablenotetext{-}{no data available} 
\end{deluxetable*}

\begin{deluxetable*}  {c c c c} 
\tablewidth{0.0pt}
\tablecaption{Column density ratios of the deuterated species $c$-$\mathrm{C_{3}HD}$, $c$-$\mathrm{C_{3}D_{2}}$ with respect to the main species in the starless  core sample. \label{tab:ratio_column_densities_pre}}
\tablehead{\colhead{Starless Core} & \colhead{$N$($c$-$\mathrm{C_{3}HD}$)/$N$($c$-$\mathrm{C_{3}H_{2}}$)} & \colhead{$N$($c$-$\mathrm{C_{3}D_{2}}$)/$N$($c$-$\mathrm{C_{3}H_{2}}$)} & \colhead{$N$($c$-$\mathrm{C_{3}D_{2}}$)/$N$($c$-$\mathrm{C_{3}HD}$)}} 
\startdata
CB23 & $0.11 \pm 0.02$ & $<0.006$ & $<0.06$ \\
L1400A & $0.08 \pm 0.02$ & \tablenotemark{-}  & \tablenotemark{-}  \\  
L1400K & $<0.03$ & \tablenotemark{-}   & \tablenotemark{-}   \\
L1495 & $0.08 \pm 0.03$ & \tablenotemark{-} & \tablenotemark{-} \\
L1495AN & $0.08 \pm 0.02$ & $0.004 \pm 0.001$  &$0.05 \pm 0.01$ \\
L1495AS & $0.08 \pm 0.02$ & \tablenotemark{-}  & \tablenotemark{-}  \\
L1495B & $0.13 \pm 0.04$ & \tablenotemark{-}  & \tablenotemark{-}  \\
L1512 & $0.08 \pm 0.02$ & $0.006 \pm 0.002$ & $0.08 \pm 0.02$ \\
L1517B & $0.10 \pm 0.03$ & $0.013 \pm 0.005$ & $0.13 \pm 0.04$ \\
TMC2 & $0.11 \pm 0.03$ & $0.012 \pm 0.003$ & $0.10 \pm 0.02$ \\
L1544 & $0.09 \pm 0.02$ & $0.015 \pm 0.003$ & $0.17 \pm 0.03$ \\
\enddata
\tablenotetext{-}{no data available} 
\end{deluxetable*}

Also in the observed protostars the deuteration for both species is similar,  as we can clearly see in Figure \ref{Fig:protostar_ratio}. The deuterium fraction peaks in HH211, reaching 23\% in case of $c$-$\mathrm{C_{3}H_2}$  and 27\% in case of $\mathrm{N_{2}H^{+}}$ which is also the highest estimated deuteration among all observed protostellar and starless cores. {The abundance ratio $N$($c$-$\mathrm{C_{3}HD}$)/$N$($c$-$\mathrm{C_{3}H_{2}}$) is within the error bars equal to 10\%, with the exception of HH211, where the average deuteration level is $0.23 \pm 0.06$. These results are similar to the $N$($c$-$\mathrm{C_{3}D_{2}}$)/$N$($c$-$\mathrm{C_{3}HD}$) ratio, that ranges from 5\% to 17\%. Finally, the abundance of the doubly deuterated species with respect to the main isotopologue varies between 0.6\% and 3.6\%.  The single and double deuteration level of $c$-$\mathrm{C_{3}H_2}$ in every protostellar core is summarized in Table \ref{tab:ratio_column_densities_pro}.

In case of IRAS16293, we derived the deuteration of $c$-$\mathrm{C_{3}H_{2}}$ and its isotopologues, as well as the depletion factor of CO by using the publicly available data from TIMASSS \citep{caux}. 
For the calculation of the $c$-$\mathrm{C_{3}H_2}$  column density and its isotopologues, we assumed a $T_{\mathrm{ex}}$  of 8.9 K for the main and the $^{13}\mathrm{C}$ species, and a $T_{\mathrm{ex}}$ of 6.3 K for the deuterated counterparts, as it was derived in \cite{majumdar}.
The $N$($c$-$\mathrm{C_{3}HD}$)/$N$($c$-$\mathrm{C_{3}H_{2}}$) ratio in IRAS16293 calculated in this work is $0.10 \pm 0.02$.  This is comparable to the deuteration of 14\% determined in \cite{majumdar} within the uncertainties.

\begin{deluxetable*}  {c c c c} 
\tablewidth{0.0pt}
\tablecaption{Column density ratios of the deuterated species $c$-$\mathrm{C_{3}HD}$, $c$-$\mathrm{C_{3}D_{2}}$ with respect to the main species in the protostellar core sample. \label{tab:ratio_column_densities_pro}}
\tablehead{\colhead{Protostellar Core} & \colhead{$N$($c$-$\mathrm{C_{3}HD}$)/$N$($c$-$\mathrm{C_{3}H_{2}}$)} & \colhead{$N$($c$-$\mathrm{C_{3}D_{2}}$)/$N$($c$-$\mathrm{C_{3}H_{2}}$)} & \colhead{$N$($c$-$\mathrm{C_{3}D_{2}}$)/$N$($c$-$\mathrm{C_{3}HD}$)}} 
\startdata
Per5 & $0.10 \pm 0.02$ & $0.017 \pm 0.005$ & $0.18 \pm 0.05$ \\
HH211 & $0.23 \pm 0.06$ & $0.036 \pm 0.009$ & $0.16 \pm 0.04$ \\  
L1448IRS2 & $0.09 \pm 0.02 $ & \tablenotemark{-}  & \tablenotemark{-}   \\
IRAS03282 & $0.14 \pm 0.05$  & \tablenotemark{-}  & \tablenotemark{-}   \\
IRAS16293 & $0.10 \pm 0.02$ & $0.012 \pm 0.003$ & $0.12 \pm 0.03$ \\
L1521F & $0.08 \pm 0.02 $ & $0.006 \pm 0.002$ & $0.08 \pm 0.02$ \\
\enddata
\tablenotetext{-}{no data available} 
\end{deluxetable*}

\subsection{Correlation between deuteration and  CO depletion factor} \label{depletion_factor}
In a cold and dense cloud, molecules in the gas phase tend to collide and freeze-out onto dust grains, leading to a gradual decrease of their gas-phase abundance. Molecules are bound on grains through Van der Waals forces \citep{garrod}, meaning that species with a non-zero dipole moment, such as CS as well as CO, will be strongly bound on grain surfaces at the low temperatures typical of starless cores; thus, they deplete from the gas phase by significant amounts. Deuteration is expected to correlate with the CO depletion factor, since CO destroys $\mathrm{H_{2}D^{+}}$ \citep{dalgarno}. Previous studies have proven that the deuterium fraction of $\mathrm{N_{2}H^{+}}$ correlates strongly with the degree of CO depletion \citep{caselli2002, crapsi1, emprecht}. The level of depletion is usually expressed as depletion factor $f_d$ and is given by:

\begin{eqnarray} \label{depletion}
f_d\mathrm{(CO)} = \frac{X_{ref}\mathrm{(CO)}}{X\mathrm{(CO)}},
\end{eqnarray} 

\noindent{where $X_{ref}\mathrm{(CO)}$ is the reference abundance of CO in the local ISM and $X\mathrm{(CO)}$ is the observed CO abundance  \citep[e.g.] [] {emprecht}}.  

Figure \ref{Fig:pre-stellar_depletion} shows the $N$($c$-$\mathrm{C_{3}HD}$)/$N$($c$-$\mathrm{C_{3}H_{2}}$) and the $N$($\mathrm{N_{2}D^{+}}$)/$N$($\mathrm{N_{2}H^{+}}$) ratio versus the CO depletion factor for the starless core sample. The depletion factors for the sources L1544, TMC2, L1495, L1517B were taken from \cite{crapsi1}.  
In order to search for a statistical correlation between $N$($c$-$\mathrm{C_{3}HD}$)/$N$($c$-$\mathrm{C_{3}H_{2}}$) and  $f_d\mathrm{(CO)}$ we have applied the Kendall's $\tau$ and the Spearman's $\rho$ rank correlation tests. The Kendall's rank test gives $\tau=0.33$ with a significance $p$ of 0.49 and the Spearman's rank test gives $\rho=0.40$ with  $p=0.60$, indicating that there is no correlation between the $c$-$\mathrm{C_{3}H_{2}}$ deuteration and the CO depletion factor within the starless core sample. 
In case of  the $\mathrm{N_{2}H^{+}}$ deuteration, however, we note a significant jump toward the pre-stellar core L1544. As already mentioned in \S \ref{results}, this indicates that $\mathrm{N_{2}H^{+}}$ is less affected by depletion than $c$-$\mathrm{C_{3}H_{2}}$ and traces the deuteration level in the highest density regions of the core; in fact the $\mathrm{N_{2}H^{+}}$ abundance also increases where CO is significantly frozen, as CO destroys $\mathrm{N_{2}H^{+}}$ to form $\mathrm{HCO^{+}}$.   

\begin{figure}[h]
	\centering
	\includegraphics[width = 0.5\textwidth]{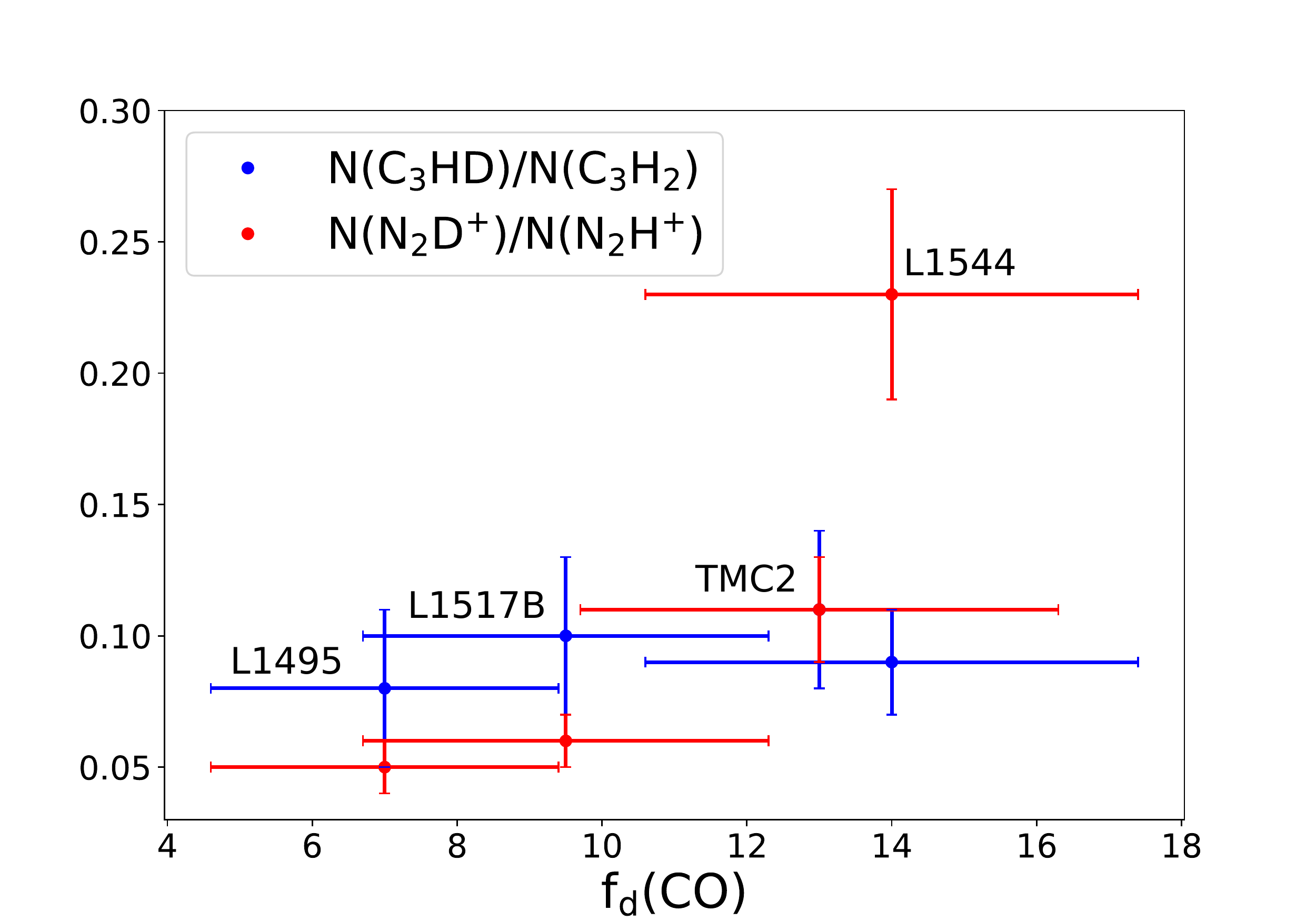}
	\caption{Deuterium fraction of $c$-$\mathrm{C_{3}H_{2}}$ and $\mathrm{N_{2}H^{+}}$ as a function of the CO depletion factor in the starless cores L1495, L1517B, TMC2 and L1544.}
	\label{Fig:pre-stellar_depletion}
\end{figure}

The correlation between $N$($c$-$\mathrm{C_{3}HD}$)/$N$($c$-$\mathrm{C_{3}H_{2}}$) and the CO depletion factor in the observed protostars is shown in Figure~\ref{Fig:protostars_depletion}. The depletion factors for the sources HH211, IRAS03282, L1448IRS2 and Per5 were taken from Emprechtinger et al. (2009), while the depletion factor for L1521F is reported in Crapsi et al. (2005). For the CO depletion level in IRAS16293 we used the spectral properties of the $\mathrm{C^{18}O}$ (1-0) transition given in \cite{caux} to calculate the column density of $\mathrm{C^{18}O}$, as it was done in \cite{emprecht}. Here we use $X_{ref}(\mathrm{C^{16}O}) = 9.5 \times 10^{-5}$ \citep{frerking} to allow fair comparison with \cite{emprecht}, although other values for $X_{ref}\mathrm{(C^{16}O)}$ can be found in the literature \citep{wannier, lacy}. We use the relation $X(\mathrm{C^{16}O})/X(\mathrm{C^{18}O}) = 560$  \citep{wilson}  such that:
\begin{eqnarray} \label{abundance}
X(\mathrm{C^{16}O)} = \frac{N_{tot}\mathrm{(C^{18}O)} \cdot 560}{N\mathrm{(H_{2})}}.
\end{eqnarray} 

The excitation temperature of  $\mathrm{C^{18}O}$ was set equal to 43 K which is the dust temperature for IRAS16293 \citep{schoier}. At this temperature no CO depletion is expected. However, the $\mathrm{C^{18}O}$ column density is an average along the line of sight, which apart from the warm central regions, also includes the cold envelope, where CO can be significantly frozen \citep [e.g.] [] {sandra}. 
The hydrogen column density $N\mathrm{(H_{2})}$ was derived by using the Herschel/SPIRE image of IRAS16293 at 250, 350 and 500 $\mu$m. These data are publicly available and can be downloaded from the Herschel Science Archive (HSA)~\footnote{www.cosmos.esa.int/web/herschel/science-archive}.  For more information concerning the data reduction, see \S \ref{density}. The resulting $N\mathrm{(H_{2})}$ of $1.0 \times 10^{23} \, \mathrm{cm^{-2}}$ is the mean value of column densities within a beam of 40$\arcsec$.
Following Equation \ref{depletion}, the depletion factor for IRAS16293 is equal to:
\begin{eqnarray} 
f_d\mathrm{(CO)} = 0.7 \pm 0.2.
\end{eqnarray} 
\noindent{This value is consistent with the low $f_d(\mathrm{CO})$ values measured by \cite{punanova} in various sources toward $\rho$ Ophiuchus.}
The fact that $f_d\mathrm{(CO)}$ is lower than 1 suggests that the adopted $X_{ref}(\mathrm{C^{16}O)}$ is underestimated by a factor of 2-3. Moreover, one has to keep in mind that the $\mathrm{C^{18}O}$ (1-0) emission was observed within 22$\arcsec$, while the estimation of $N\mathrm{(H_{2})}$ was done within a 40$\arcsec$ beam. This indicates that the derived depletion factor should be considered as a lower limit.

Figure \ref{Fig:protostars_depletion} shows no clear trend between the deuteration of $c$-$\mathrm{C_{3}H_{2}}$ and the CO depletion among the protostellar cores. This is also confirmed by the Kendall's rank test that gives $\tau = 0.32$ with a significance $p$ of 0.45 as well as by the Spearman's rank test that results to $\rho=0.56$ with $p=0.32$ (without considering the protostar L1521F).
The source L1521F deviates strongly from the rest of the sources, having a significant depletion factor of 15 and simultaneously showing a low  $c$-$\mathrm{C_{3}H_{2}}$ deuteration of 8\%. These values are comparable to those found in L1544, where the CO depletion factor is 14 \citep{crapsi1} and the deuteration of $c$-$\mathrm{C_{3}H_{2}}$ is 9\%. The peculiarity of L1521F has been proven already in previous works, where the central column density $N\mathrm{(H_{2})}$ is high ($13.5  \times 10^{22} \, \mathrm{cm^{-2}}$) despite the low $\mathrm{N_{2}H^{+}}$ deuteration \citep [being a factor of 2 lower than in  L1544;] [] {crapsi2, crapsi1}. 

The large CO depletion factor and low deuteration both in $\mathrm{N_{2}H^{+}}$ and $c$-$\mathrm{C_{3}H_{2}}$ in L1521F, could be a signature of episodic accretion of the central protostar. Since L1521F has been classified as a VeLLO \citep{bourke, Takahashi} it may be in a quiescent phase, following an accretion burst event. During such a burst, CO is expected to return in the gas phase \citep{Visser}, thus reducing the deuterium fraction. After the burst, the fast cooling of the dust could quickly lead to CO freeze out, with short time scales of the order of $10^9/\mathrm{n_H \, yr}$, where $n_{\mathrm{H}}$ is the total number density of hydrogen nuclei \citep[e.g.] [] {caselli99}, while the deuteration of gas-phase molecules is a slower process, especially if during the burst the ortho-to-para $\mathrm{H_2}$ ratio (sensitive to the temperature) increases to values larger than 1\% \citep [e.g.] [] {flower, kong}.
However, the exact physical and chemical conditons of L1521F are beyond the scope of this work and it is clear that further observations are needed to prove this point.

\begin{figure}[h]
	\centering
	\includegraphics[width = 0.5\textwidth]{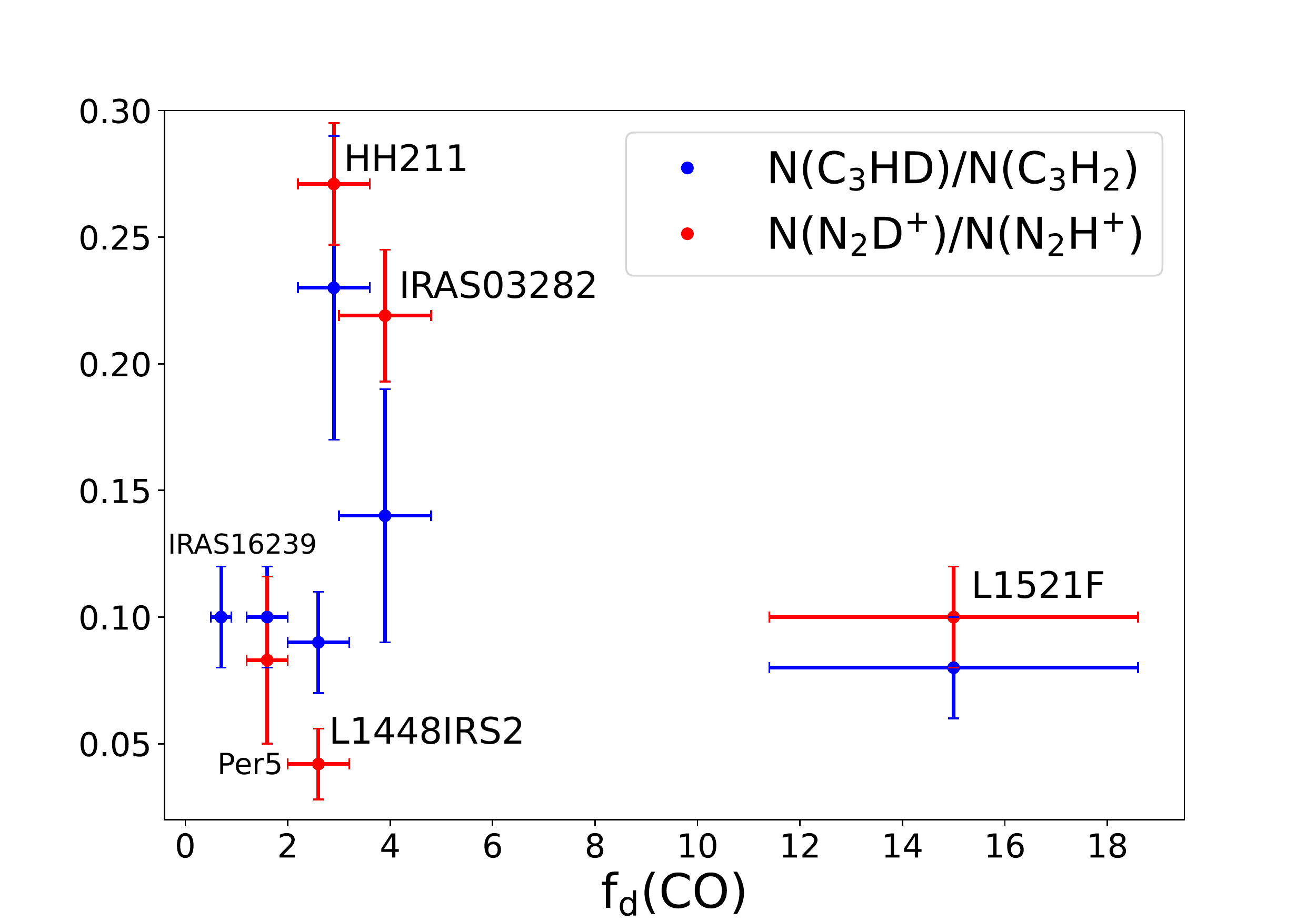}
	\caption{Deuterium fraction of $c$-$\mathrm{C_{3}H_{2}}$ and $\mathrm{N_{2}H^{+}}$ as a function of CO depletion factor for the protostellar core sample. The deuterium fraction of $\mathrm{N_{2}H^{+}}$ and the CO depletion factors were taken from \cite{emprecht}, while the $N$($\mathrm{N_{2}D^{+}}$)/$N$($\mathrm{N_{2}H^{+}}$) ratio and   $f_d\mathrm{(CO)}$ for L1521F was given in \cite{crapsi1}. }
	\label{Fig:protostars_depletion}
\end{figure}

\subsection{Correlation between deuteration and $\mathrm{H_{2}}$ column density} \label{density}
As already pointed out in \S \ref{depletion_factor}, an increase in volume density toward the core centre is expected to correlate with an increase in deuteration, because of the consequently larger CO freeze-out rates, as long as the temperature remains below 20 K. For this reason we examine how the estimated deuteration of $c$-$\mathrm{C_{3}H_{2}}$ in the starless and the pre-stellar cores correlates with the central column density of molecular hydrogen. 
For the $N\mathrm{(H_{2})}$ calculation we use the Herschel/SPIRE image of the Taurus Complex at 250, 350 and 500 $\mu$m, taken during the Herschel Gould Belt Survey \citep{andre}. 

\begin{deluxetable*} {c c c} 
\tablewidth{0.0pt}
\tablecaption{$N\mathrm{(H_{2})}$ values for the starless core sample, derived from the Herschel/SPIRE images, within a 40$\arcsec$ beam (this work), as well as the ones derived from the 1.2 mm continuum emission within a 11$\arcsec$ beam \citep{crapsi1}. \label{tab:herschel}}
\tablehead{\colhead{Starless Core} &  \colhead{$N\mathrm{(H_{2})}$\tablenotemark{a}}  &  \colhead{$N\mathrm{(H_{2})}$\tablenotemark{b}}\\
&\colhead{ ($10^{21}$ $\mathrm{cm^{-2}}$)} & \colhead{($10^{21}$ $\mathrm{cm^{-2}}$)}} 
\startdata
L1495 & $8.3 \pm 1.6$ & $31 \pm 10$ \\
L1495B & $10.1 \pm 2.0$  \\  
L1495AN & $15.1 \pm 2.7$  \\
L1495AS & $22.1 \pm 4.1$ \\
TMC2 & $18.9 \pm 3.3$  & $60 \pm 12$\\
L1544 & $27.9 \pm 5.1 $ & $ 94 \pm 16$   \\
L1512 & $8.6 \pm 1.6 $   \\
L1517B & $11.5 \pm 2.3 $  & $37 \pm 10$ \\
CB23 & $8.5 \pm 1.6 $   \\
\enddata
\tablenotetext{a}{ This work}
\tablenotetext{b}{\cite{crapsi1}}
\end{deluxetable*}

A modified blackbody radiation is fitted to each pixel, using an emissivity spectral index $\beta=1.5$  and a dust emissivity coefficient $\kappa_{250 \mathrm{\mu m}} = 0.1 \, \mathrm{g^{-1} \, cm^{2}}$ \citep{hilde}. The data reduction involves smoothing the 250 $\mu$m and 350 $\mu$m images to the resolution of the 500 $\mu$m image and resampling all images to the same grid. From this fitting procedure we obtain the central column density $N\mathrm{(H_{2})}$ as well as the dust temperature $T_{\mathrm{Dust}}$ for every pixel. The resulting  $N\mathrm{(H_{2})}$ is the mean value of column densities in a 40$''$ beam. For the error estimation we take into account calibration uncertainties of 7\% according to the SPIRE manual (see Appendix \ref{error_est}). In Table \ref{tab:herschel} we summarize the resulting column densities and their uncertainties. The estimated $N\mathrm{(H_{2})}$ values are up to a factor of 3 smaller than the ones calculated in \cite{crapsi1} at the $1.2 \,\mathrm{mm}$ dust continuum peak. This results from the different beam sizes used, 40$\arcsec$ vs 11$\arcsec$, so that the highest density parts of the central cores \citep [such as L1544, with central densities of $\sim10^6 \, \mathrm{cm^{-3}}$ within a 1000 AU in radius; see] [] {keto} are diluted within the $Herschel$ beam. \\ 
Figure \ref{Fig:prestellar_central_density} shows the correlation between $N$($c$-$\mathrm{C_{3}HD}$)/$N$($c$-$\mathrm{C_{3}H_{2}}$)  and $N\mathrm{(H_{2})}$ for the starless core sample. We also included the $N$($\mathrm{N_{2}D^{+}}$)/$N$($\mathrm{N_{2}H^{+}}$) ratio, calculated in \cite{crapsi1} for a direct comparison. The cores L1400K and L1400A were not part of this study, since there were no SPIRE images of these sources available.

\begin{figure}[h]
	\centering
	\includegraphics[width = 0.5\textwidth]{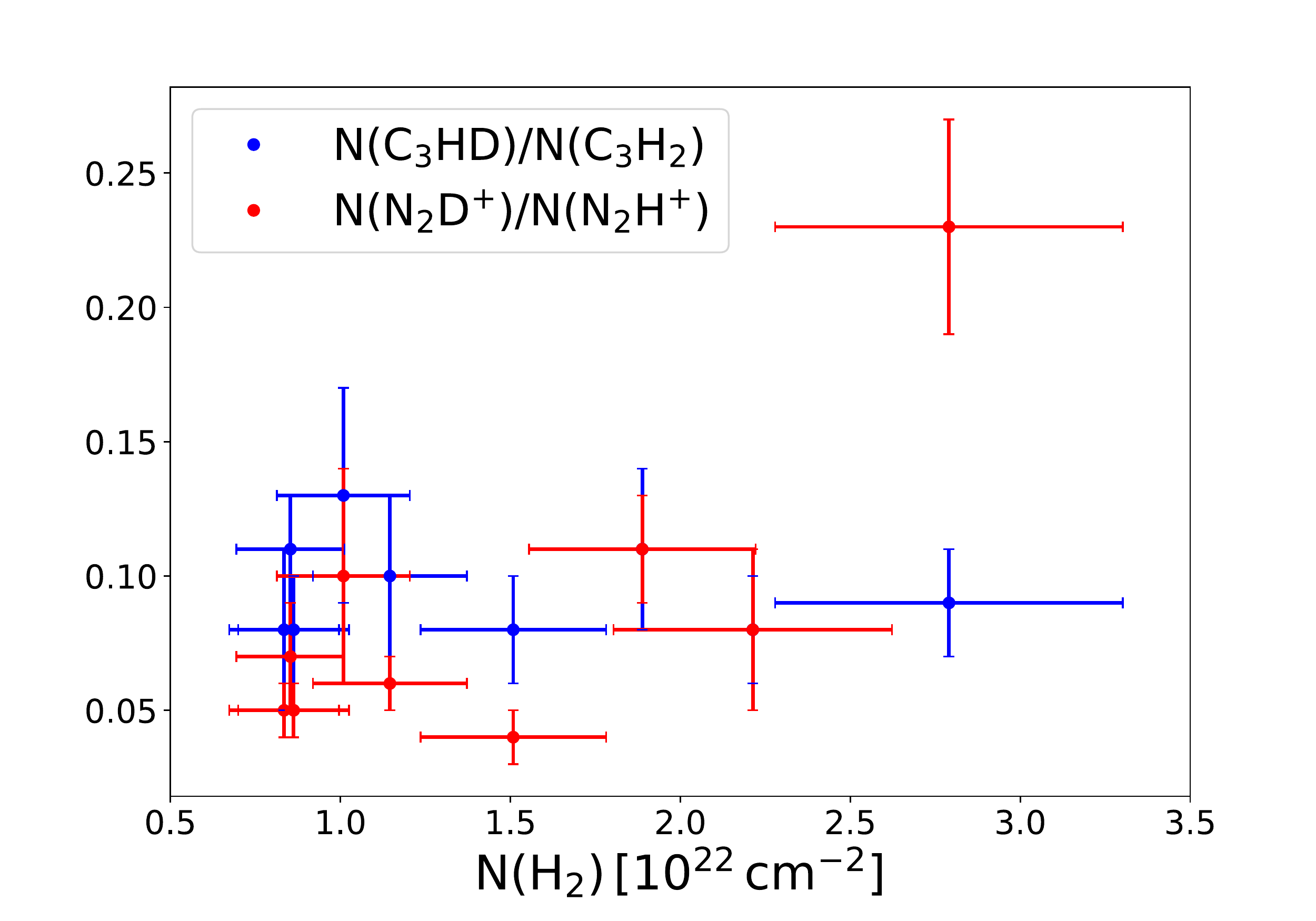}
	\caption{Deuterium fraction of $c$-$\mathrm{C_{3}H_{2}}$ (blue dots) and $\mathrm{N_{2}H^{+}}$ (red dots) as a function of central column density of $\mathrm{H_{2}}$, measured with $Herschel$ in the starless and pre-stellar core sample. The deuteration of $\mathrm{N_{2}H^{+}}$  was calculated in \cite{crapsi1}.}
	\label{Fig:prestellar_central_density}
\end{figure}

No correlation is found between the deuteration level of both species and $N\mathrm{(H_{2})}$ among the starless cores. However, in L1544, we recognize a substantial increase in the deuteration of $\mathrm{N_{2}H^{+}}$, as it was already visible in Figure \ref{Fig:pre-stellar_ratio}. 
 In \cite{crapsi1} there are three additional evolved pre-stellar cores (L183, L429 and L694-2), that show enhanced  $\mathrm{N_{2}H^{+}}$ deuteration with increasing $N(\mathrm{H_2})$. This indicates that indeed $\mathrm{N_{2}H^{+}}$ is a late-type molecule and stays in the gas phase at high densities, while  $c$-$\mathrm{C_{3}H_{2}}$ is possibly depleted in the central regions, thus it stops tracing the central zone of the core where high levels of deuterium fractions are present. This is in agreement with our current understanding of the chemistry of  \mbox{$c$-$\mathrm{C_{3}H_{2}}$}, and its distribution across the pre-stellar core L1544 \citep[see Figure \ref{Fig:model_sipila} in Appendix \ref{sipila}] {sipila16}. 

Another way of testing this theory is to examine parameters that are related to the kinematic of the gas, such as the width of the detected lines. 
Figure \ref{Fig:prestellar_linewidth} shows the correlation between the observed linewidth, $\Delta \varv_{\mathrm{obs}}$, of $c$-$\mathrm{C_{3}H_{2}}$ ($3_{2,2}-3_{1,3}$) and of $\mathrm{N_{2}H^{+}}$ (1-0) among the starless and protostellar core sample. Thermal, turbulent and systematic motions contribute to the total $\Delta \varv_{\mathrm{obs}}$. Thermal broadening does not play a substantial role, since the thermal linewidth of $c$-$\mathrm{C_{3}H_{2}}$ is just 0.11 $\mathrm{km \, s^{-1}}$ at 10 K (and 0.13 $\mathrm{km \, s^{-1}}$ for $\mathrm{N_{2}H^{+}}$). As we can clearly see in Figure \ref{Fig:prestellar_linewidth}, the observed $c$-$\mathrm{C_{3}H_{2}}$ line has a larger width in most of the cores (except for L1495 and L1495AN, where the observed linewidths are approximately the same).  

\begin{figure}[h]
	\centering
	\includegraphics[width = 0.5\textwidth]{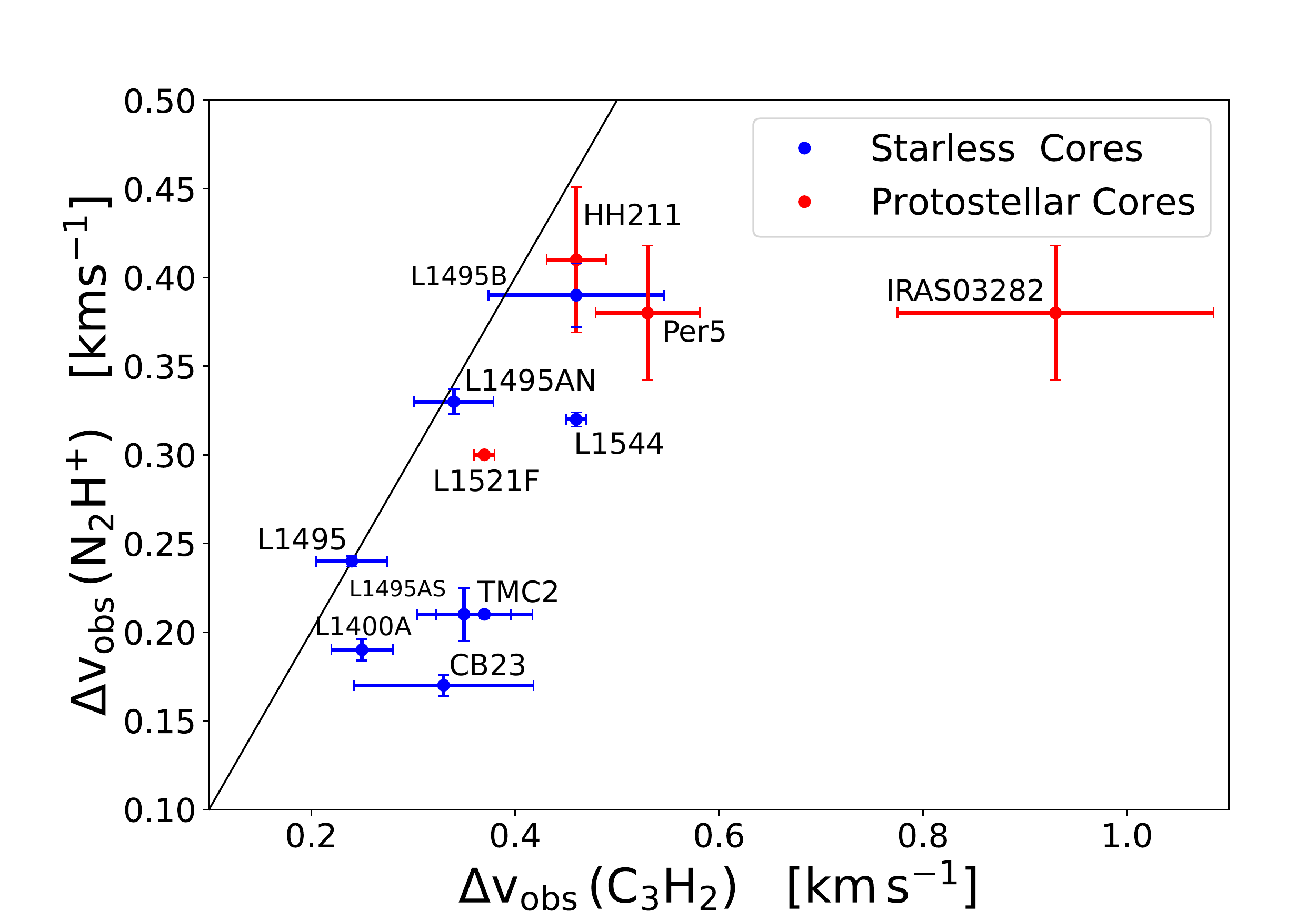}
	\caption{Observed linewidth of $\mathrm{N_{2}H^{+}}$ \citep{crapsi1, emprecht} compared to that of $c$-$\mathrm{C_{3}H_{2}}$ (this work) in the starless and protostellar core sample. The dashed line represents the 1:1 ratio.} 
	\label{Fig:prestellar_linewidth}
\end{figure}

This suggests that $c$-$\mathrm{C_{3}H_{2}}$ traces a different region than $\mathrm{N_{2}H^{+}}$, where turbulent or systematic (i.e. rotation, infall or outflows in protostellar objects) motion dominates. In both starless and protostellar core samples we see the same trend by studying the linewidth of $c$-$\mathrm{{H^{13}CC_{2}H}}$  ($2_{1,2}-1_{0,1}$) instead, suggesting that the optical depth broadening of $c$-$\mathrm{C_{3}H_{2}}$ ($3_{2,2}-3_{1,3}$) is negligible.

\subsection{Correlation between deuteration and dust temperature} \label{temp}
During the protostellar phase, the core starts to warm up the surrounding material leading to desorption of CO from dust grains. Moreover, the energetic ouflows by protostars can also contribute to the release of CO molecules in the gas phase via sputtering or grain-grain collisions \citep[e.g.] [] {caselli97,jimenez}. Back in the gas phase, CO can react and destroy $\mathrm{H_{2}D^{+}}$, reducing the total deuteration in molecules. In addition,  reaction \ref{deuteration} can proceed at high temperatures also in the backward direction, leading to a further reduction of $\mathrm{H_{2}D^{+}}$. Therefore, one expects an anticorrelation between the deuteration and the dust temperature \citep[e.g.] [] {ladd, myers} of a protostar, as it has already been confirmed in previous works \citep{emprecht, fontani}. In Figure \ref{Fig:protostars_tdust_L1521F} we plot the deuterium fraction of \mbox{$c$-$\mathrm{C_{3}H_{2}}$} against $T_{\mathrm{Dust}}$ in the protostellar core sample. The values for the $T_{\mathrm{Dust}}$ were taken from \cite{emprecht}. As in \S \ref{depletion_factor}, also here we include the protostar IRAS16293  \citep{schoier}.  
Figure \ref{Fig:protostars_tdust_L1521F} shows a clear anticorrelation between deuteration and $T_{\mathrm{Dust}}$, when excluding L1521F, that has a Kendall's $\tau$ coefficient of -0.74 with a significance $p$ of 0.08 and a Spearman's $\rho$ coefficient of -0.82 with a significance $p$ of 0.09.
The abundance ratio $N$($c$-$\mathrm{C_{3}HD}$)/$N$($c$-$\mathrm{C_{3}H_{2}}$) peaks in the coldest source, HH211 at 23\%, and decreases toward the warmer sources down to 8\%.

The young source HH211 could be an example of a protostar where the accretion burst has recently happened and/or where ices have been recently evaporated. If this is the case, one way to interpret the large deuteration of $c$-$\mathrm{C_{3}H_{2}}$ is the release of large amount of deuterated (and non-deuterated) $c$-$\mathrm{C_{3}H_{2}}$ from the ices into the gas phase.  Furthermore, this could imply significant deuteration of $c$-$\mathrm{C_{3}H_{2}}$ on the surface of dust grains, maybe due to hydrogen-deuterium exchange reactions known to happen for other molecules, such as $\mathrm{CH_3OH}$ during the preceding cold pre-stellar phase \citep[e.g.]{parise06}. \footnote{Recent experiments of D-H exchanges  carried out by \cite{faure} suggest that D-H exchanges are made possible by hydrogen bonds between the hydroxyl functional groups of methanol and water ice. This makes it unlikely that such a process could work for $c$-$\mathrm{C_{3}H_{2}}$, although experimental and theoretical work is needed to rule out this hypothesis.} 

The result for L1521F deviates considerably from the rest of the sources. The dust temperature for L1521F  ($T_{\mathrm{Dust}} = 9 \pm 2 \, \mathrm{K}$) was taken from \cite{kirk}. As already highlighted in \S \ref{depletion_factor}, L1521F hosts a VeLLO which could be a protostar in very early stages of evolution and/or an example of low activity in protostellar evolution characterized by episodic accretion. This could explain the mismatch between the physical (low $T_{\mathrm{Dust}}$) and chemical conditions (low deuteration).

\begin{figure}[! h]
	\centering
	\includegraphics[width = 0.5\textwidth]{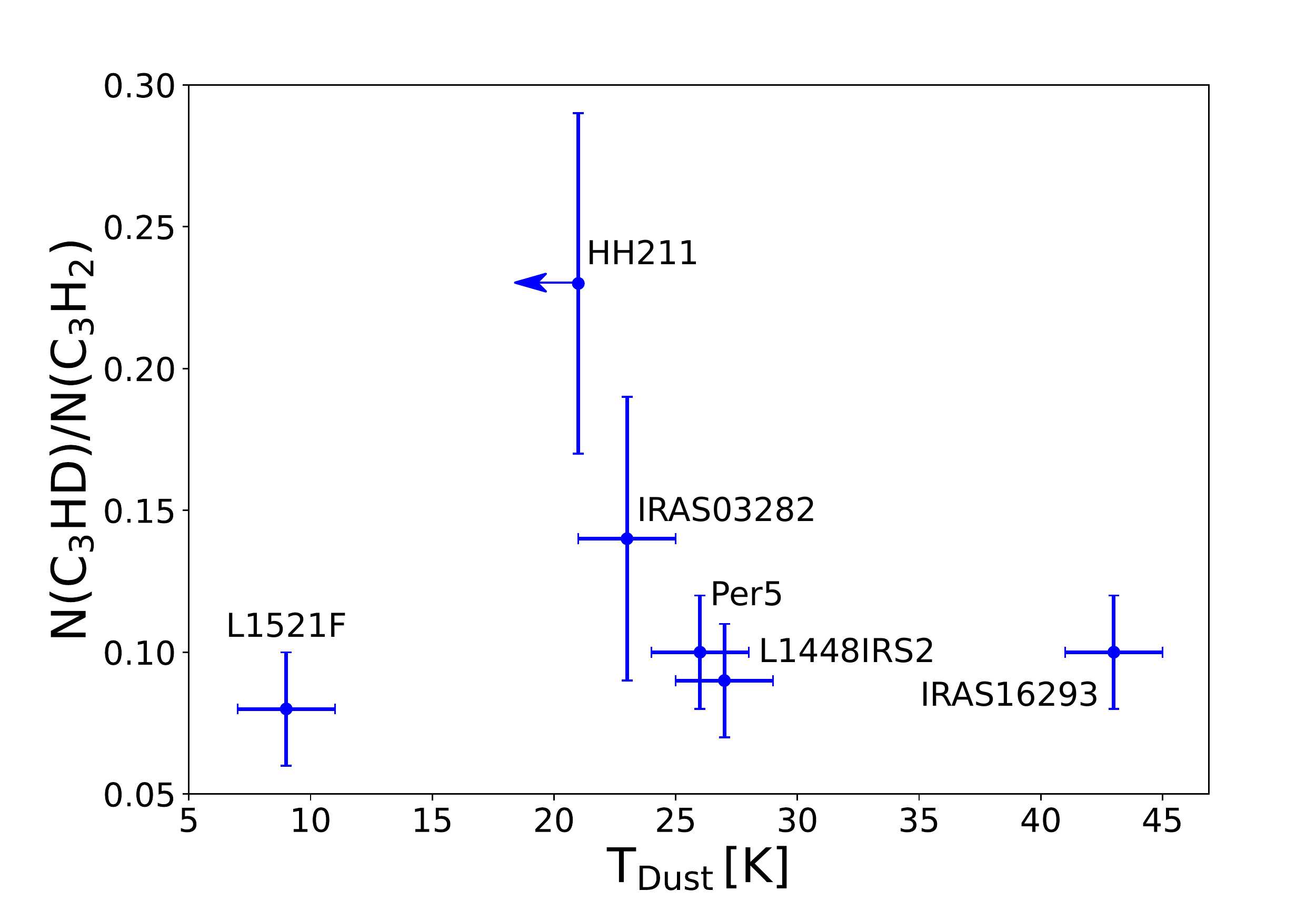}
	\caption{Deuterium fraction of $c$-$\mathrm{C_{3}H_{2}}$ as a function of dust temperature $T_{\mathrm{Dust}}$ in the protostellar core sample, including the VeLLO L1521F.}
	\label{Fig:protostars_tdust_L1521F}
\end{figure}

\section{Conclusion} \label{con}
In this work we present single pointing observations of $c$-$\mathrm{C_{3}H_{2}}$, the singly and doubly deuterated species, $c$-$\mathrm{C_{3}HD}$ and $c$-$\mathrm{C_{3}D_{2}}$ as well as the isotopologue $c$-$\mathrm{{H^{13}CC_{2}H}}$ toward 10 starless cores and 5 protostellar cores in the Taurus and the Perseus Complexes. The  pre-stellar core L1544 and the protostellar core IRAS16293 were also included in our study. We calculated the deuterium fractionation $N$($c$-$\mathrm{C_{3}HD}$)/$N$($c$-$\mathrm{C_{3}H_{2}}$) and studied its correlation with the CO depletion factor, the dust temperature and the $\mathrm{H_2}$  column density toward the core centre. We also examined the differences between the deuteration of C- and N-bearing molecules. Here is a summary of our main conclusions:  \\
\begin{itemize}
\item The ratio $N$($c$-$\mathrm{C_{3}HD}$)/$N$($c$-$\mathrm{C_{3}H_{2}}$) is within the error bars equal to 10\% in all starless and pre-stellar cores, where $c$-$\mathrm{C_{3}HD}$ has been detected. This also accounts for the protostars except for the source HH211 where we measure a high deuteration level of $(23 \pm 6)\%$. The  \mbox{$N$($c$-$\mathrm{C_{3}D_{2}}$)/$N$($c$-$\mathrm{C_{3}H_{2}}$)} ratio ranges from 0.4 to 1.5\% in starless and pre-stellar cores, and from 0.6 to 3.6\% among the protostellar cores.
The deuteration of $c$-$\mathrm{C_{3}H_{2}}$ and $\mathrm{N_{2}H^{+}}$
follows the same trend in both samples. However, in case of the evolved pre-stellar core L1544, the deuteration of $\mathrm{N_{2}H^{+}}$ is significantly higher (factor of 2.6) compared to the $N$($c$-$\mathrm{C_{3}HD}$)/$N$($c$-$\mathrm{C_{3}H_{2}}$) ratio. This can be understood by the well-known fact that $\mathrm{N_{2}H^{+}}$ remains in the gas phase even at densities of about $10^6 \, \mathrm{cm^{-3}}$, where the C-bearing molecules (including $c$-$\mathrm{C_{3}H_{2}}$ and its deuterated forms) are highly frozen and where deuterium fractionation proceeds fast, thanks to the large abundances of the deuterated $\mathrm{H_3^{+}}$ isotopologues.  \\
\item Among the starless cores, we find no correlation between the deuteration of $c$-$\mathrm{C_{3}H_{2}}$  and the CO depletion factor as well as the central $\mathrm{H_{2}}$ column density (measured with $Herschel$). However, the $\mathrm{N_{2}H^{+}}$ deuteration substantially increases toward L1544 which is the most evolved source within the starless core sample. This indicates that $\mathrm{N_{2}H^{+}}$ traces the central dense core, unlike $c$-$\mathrm{C_{3}H_{2}}$, which is more likely tracing an outer shell, surrounding the dense region of the pre-stellar core. This theory is also favoured by the fact that the observed linewidth of the $c$-$\mathrm{C_{3}H_{2}}$ emission is larger than that of the $\mathrm{N_{2}H^{+}}$ emission among the starless and the protostellar cores. As a consequence, the information resulting from observing $c$-$\mathrm{C_{3}H_{2}}$ and $\mathrm{N_{2}H^{+}}$ is complementary because they bring insights on different regions of the core.\\
\item Among the protostellar cores, there is a tight anticorrelation between the deuteration of $c$-$\mathrm{C_{3}H_{2}}$ and the dust temperature ($\tau=-0.74$ and $\rho=-0.82$ ), with the exception of L1521F. The $N$($c$-$\mathrm{C_{3}HD}$)/$N$($c$-$\mathrm{C_{3}H_{2}}$) ratio drops with increasing $T_{\mathrm{Dust}}$, reaching the maximum value of 0.23 in the coolest source HH211 ($T_{\mathrm{Dust}} < 21$~K) and decreasing toward the warmest sources, down to a deuteration level of 0.08. The $c$-$\mathrm{C_{3}H_{2}}$ deuteration does not correlate with the CO depletion factor within the protostellar core sample. L1521F differs substantially from the rest of the sources, showing in both species, $c$-$\mathrm{C_{3}H_{2}}$ and $\mathrm{N_{2}H^{+}}$, a low deuteration and at the same time a high depletion factor of 15 (see comment below for explanation).\\
\item The high $c$-$\mathrm{C_{3}H_{2}}$ deuteration of the youngest source HH211, might be the result of a recent evaporation of $c$-$\mathrm{C_{3}H_{2}}$ and $c$-$\mathrm{C_{3}HD}$, coming from a recent accretion burst. The timescale must be short enough to avoid the enhanced abundance of CO in the gas phase to significantly alter the $c$-$\mathrm{C_{3}H_{2}}$ abundance and deuterium fraction. As similarly large values are not seen in the pre-stellar phase, deuteration processes taking place also on dust grains would be a possible scenario for the observed high deuteration of $c$-$\mathrm{C_{3}H_{2}}$ in this young protostar. \\
\item The source L1521F shows a peculiar behaviour, having a low deuteration of 0.08 in spite of a significant CO depletion factor ($f_d\mathrm{(CO)} = 15 \pm 3.6$) and a low dust temperature ($T_{\mathrm{Dust}} =9 \pm 2 \, \mathrm{K}$). The peculiarity of this source has been highlighted in other studies as well  \citep{crapsi1, crapsi2}. L1521F could be an episodically accreting low-mass protostar in a quiescent phase \citep{bourke, Takahashi}. The fact that it shows a low deuteration and dust temperature, while having a high CO depletion, might imply that after a recent burst, which heated dust and gas,  the ortho-to-para-$\mathrm{H_2}$ ratio possibly increased, thus slowing down the deuteration process in the now cold envelope. 
\end{itemize}

\acknowledgements{We thank the anonymous Referee for his/her comments that significantly improved the present work. S. Spezzano acknowledges the financial support of the Minerva Fast Track fellowship of the Max Planck Gesellschaft.}

\appendix

\section{Observed Spectra  of $c$-$\mathrm{C_{3}H_{2}}$ and its isotopologues toward the starless and protostellar core samples } \label{spectra}

\begin{figure*}[h]
	\centering
	\includegraphics[width = 1\textwidth]{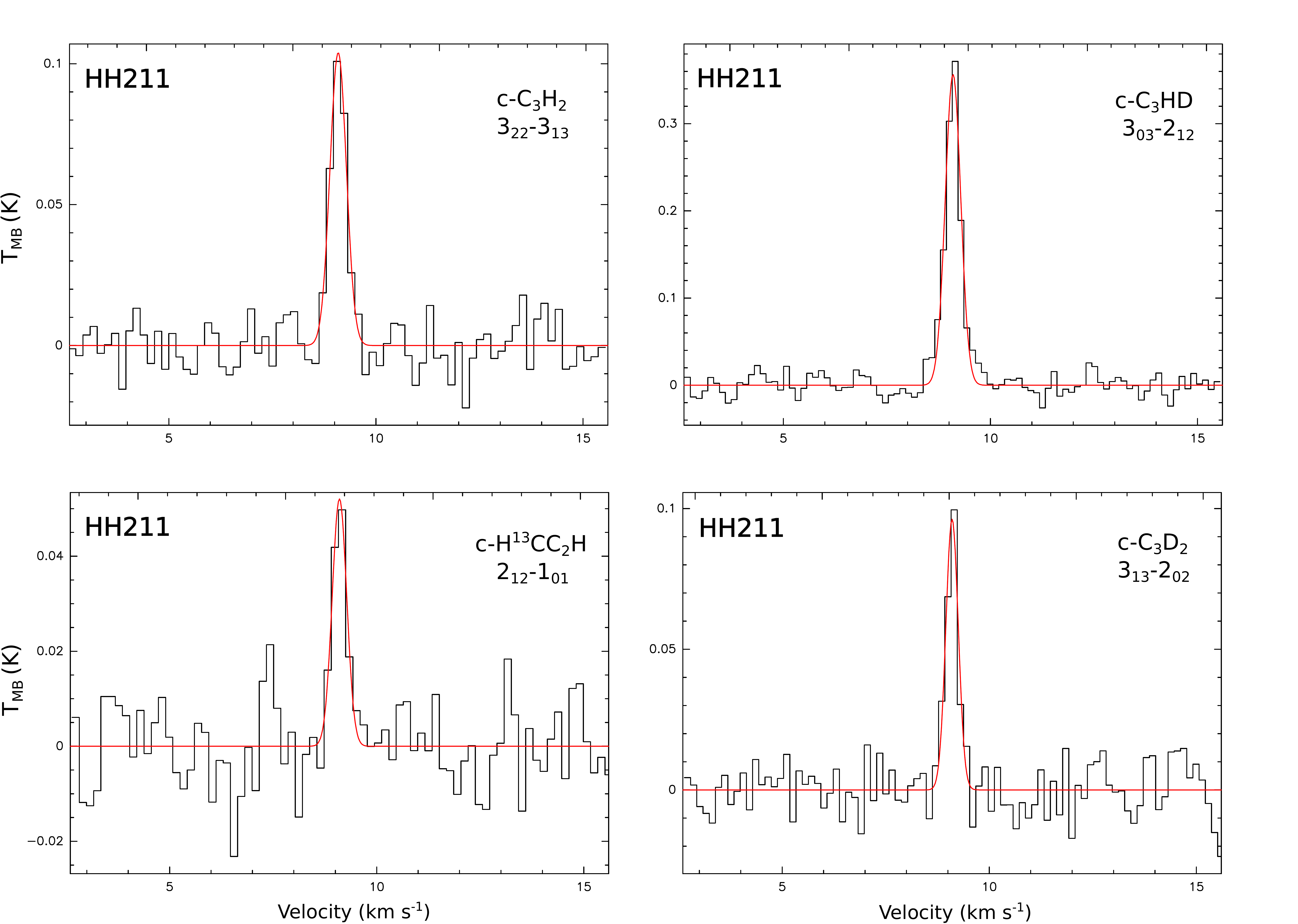}
	\caption{Spectra of several isotopologues of $c$-$\mathrm{C_{3}H_{2}}$ toward the protostellar core HH211. The red line plots the CLASS Gaussian fit.}
	\label{Fig:ratios1}
\end{figure*}

\begin{figure*}[h]
	\centering
	\includegraphics[width = 1\textwidth]{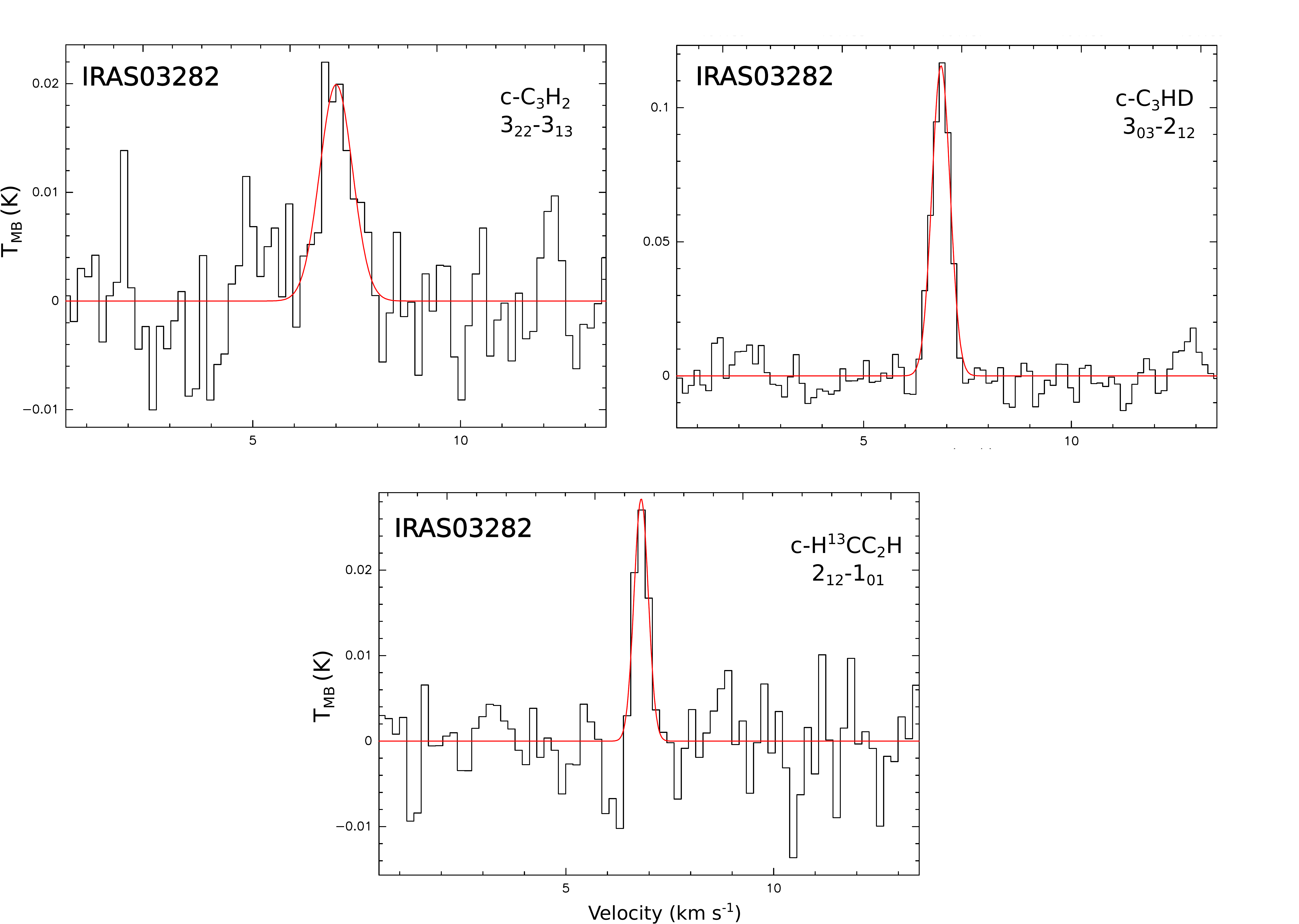}
	\caption{Spectra of the isotopologues of $c$-$\mathrm{C_{3}H_{2}}$  observed toward the protostellar core IRAS03282. The red line plots the CLASS Gaussian fit.}
	\label{Fig:ratios2}
\end{figure*}

\newpage

\begin{figure*}[h]
	\centering
	\includegraphics[width = 1\textwidth]{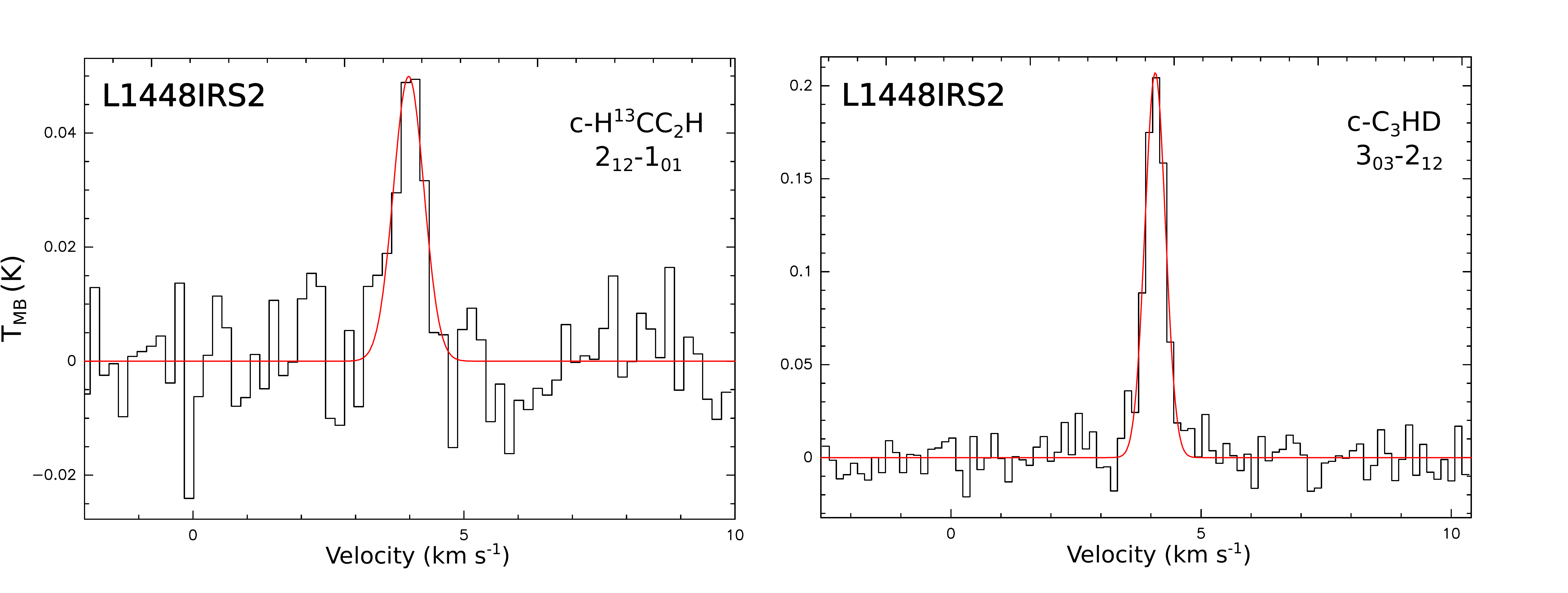}
	\caption{Spectra of the isotopologues of $c$-$\mathrm{C_{3}H_{2}}$ observed toward the protostellar core L1448IRS2. The red line plots the CLASS Gaussian fit.}
	\label{Fig:ratios3}
\end{figure*}

\begin{figure*}[h]
	\centering
	\includegraphics[width = 1\textwidth]{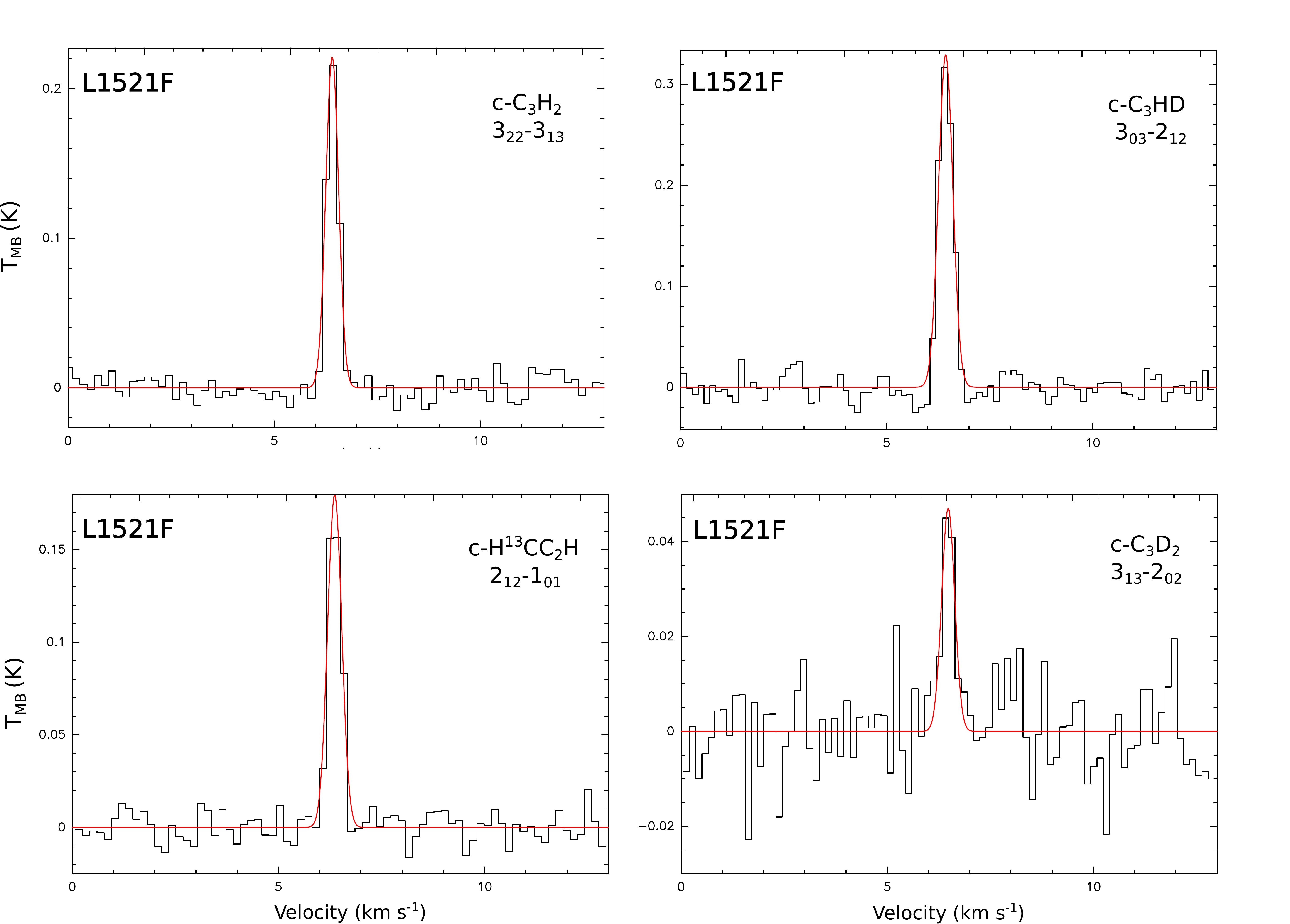}
	\caption{Spectra of the isotopologues of $c$-$\mathrm{C_{3}H_{2}}$ observed toward the protostellar core L1521F. The red line plots the CLASS Gaussian fit.}
	\label{Fig:ratios4}
\end{figure*}

\begin{figure*}[h]
	\centering
	\includegraphics[width = 1\textwidth]{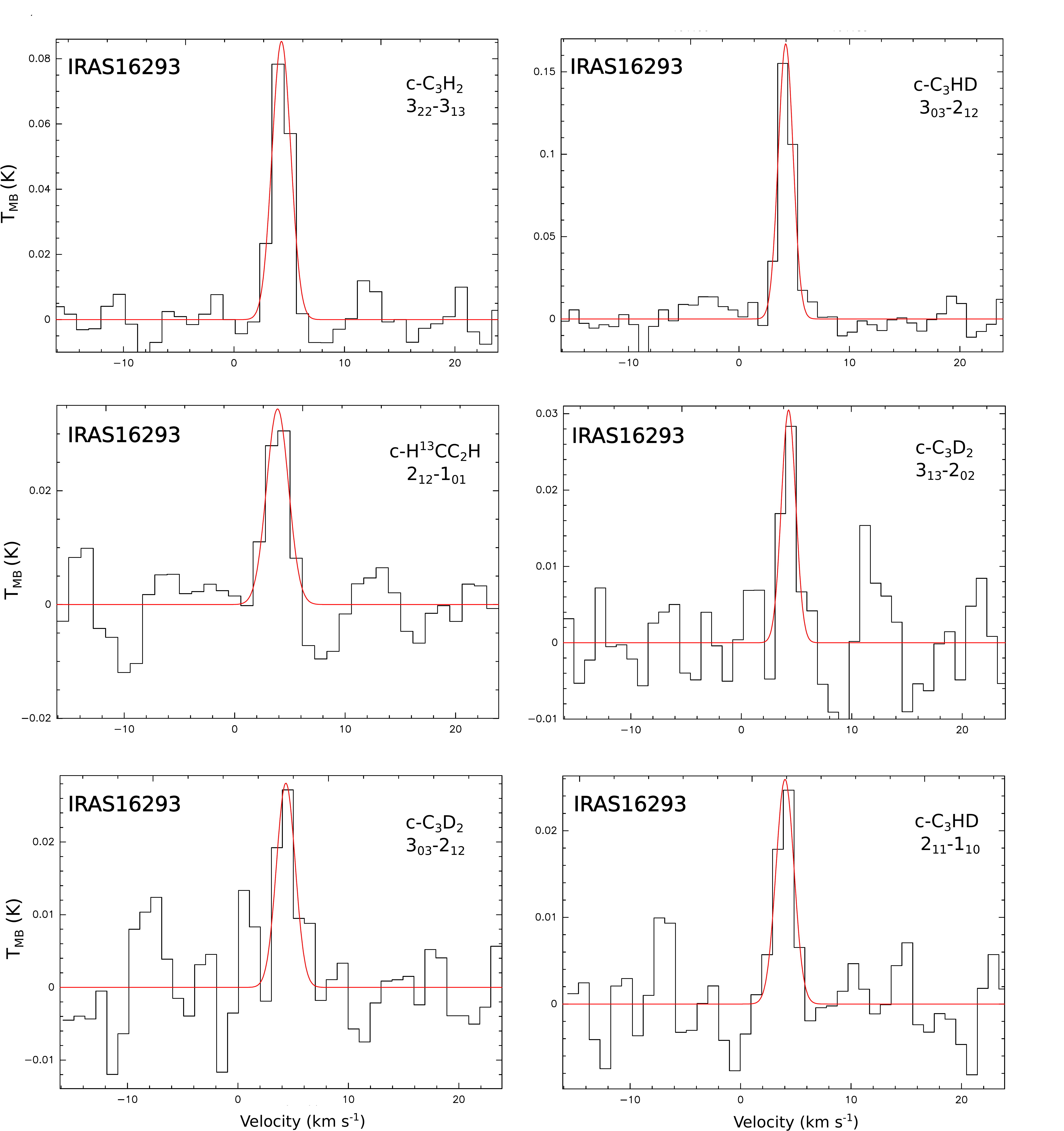}
	\caption{Spectra of the isotopologues of $c$-$\mathrm{C_{3}H_{2}}$ observed toward the protostellar core IRAS16293. The red line plots the CLASS Gaussian fit.}
	\label{Fig:ratios5}
\end{figure*}

\begin{figure*}[h]
	\centering
	\includegraphics[width = 1\textwidth]{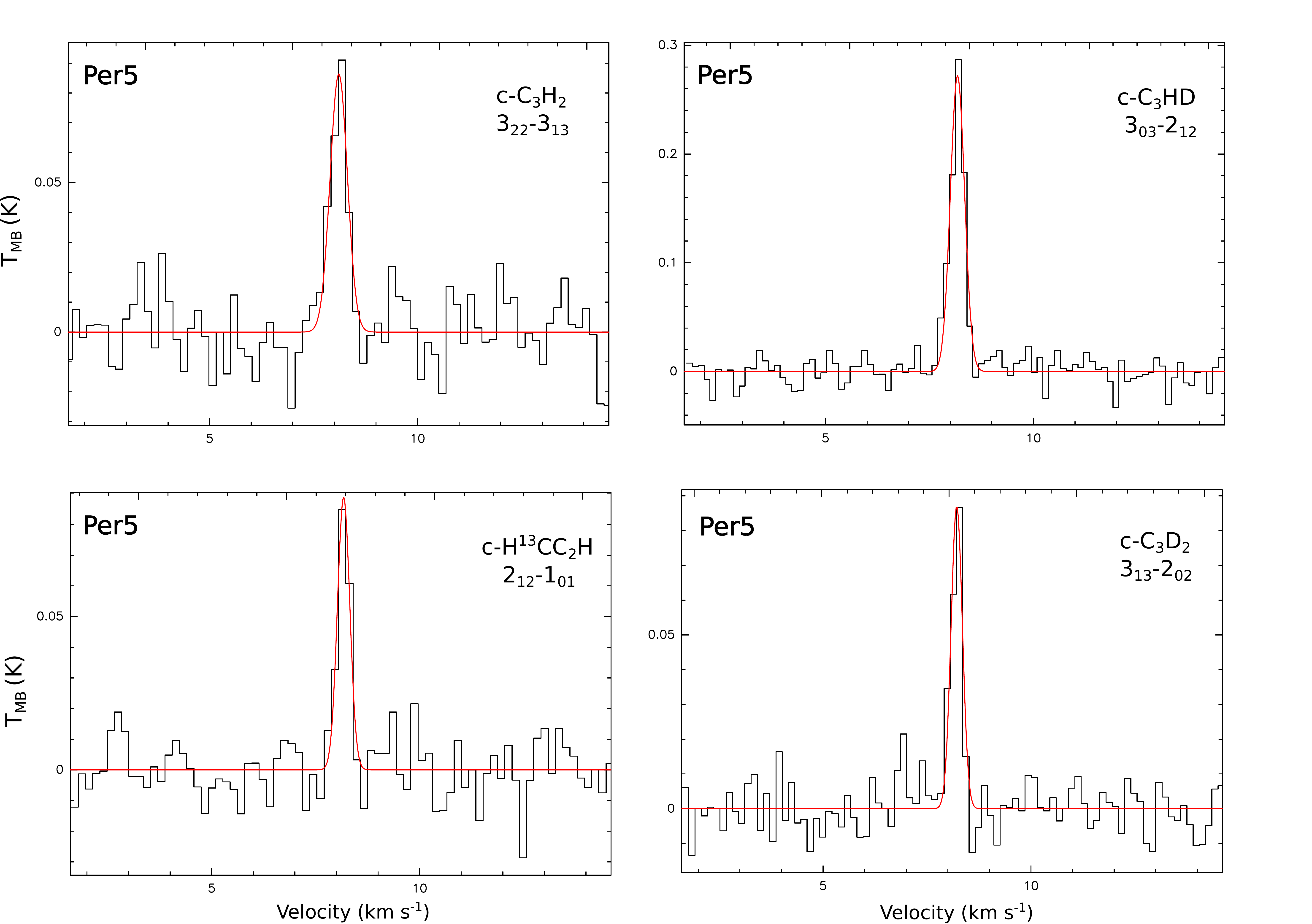}
	\caption{Spectra of the isotopologues of $c$-$\mathrm{C_{3}H_{2}}$ observed  toward the protostellar core Per5. The red line plots the CLASS Gaussian fit.}
	\label{Fig:ratios15}
\end{figure*}

\begin{figure*}[h]
	\centering
	\includegraphics[width = 1\textwidth]{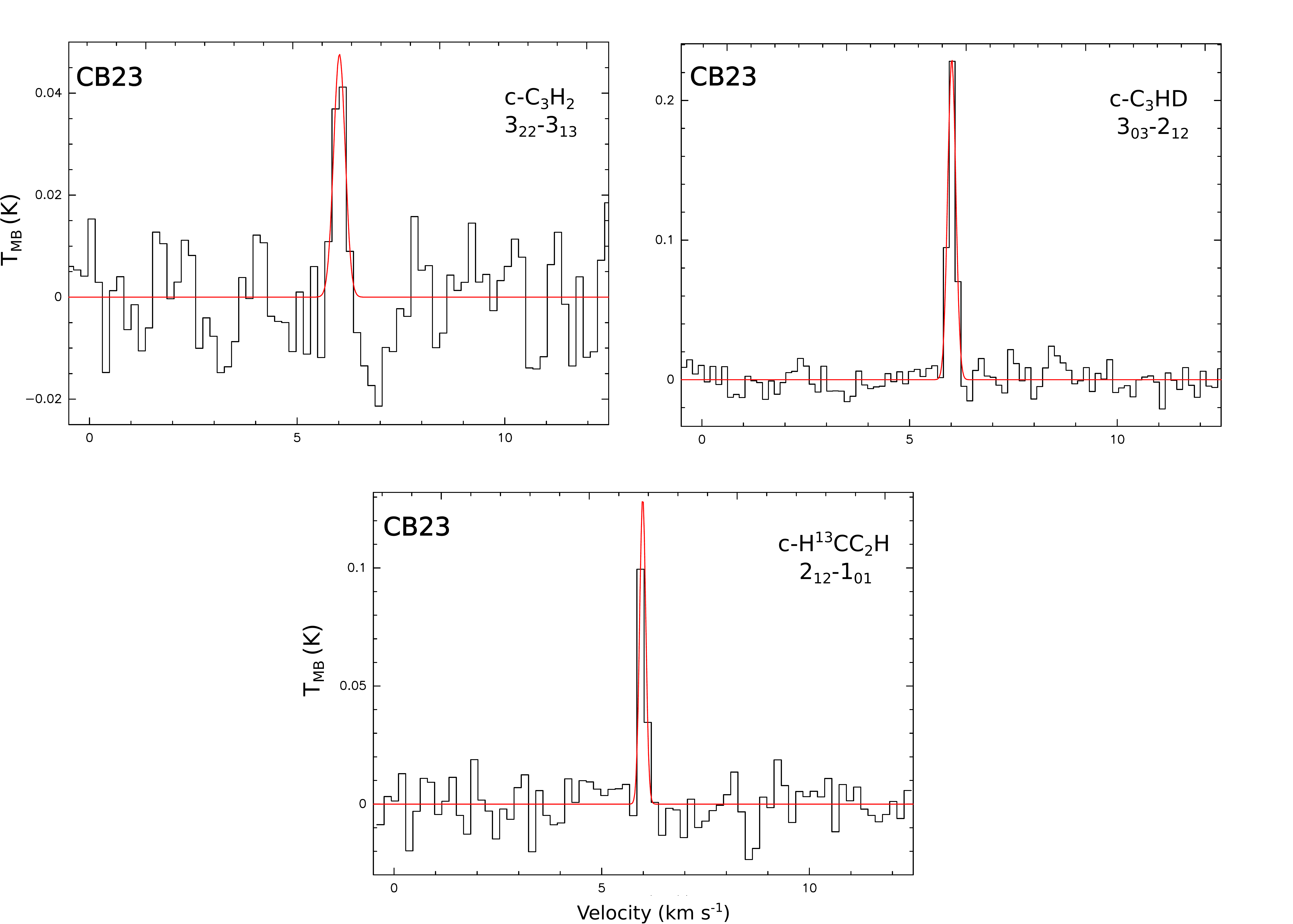}
	\caption{Spectra of the isotopologues of $c$-$\mathrm{C_{3}H_{2}}$ observed  toward the starless core CB23. The red line plots the CLASS Gaussian fit.}
	\label{Fig:ratios6}
\end{figure*}

\begin{figure*}[h]
	\centering
	\includegraphics[width = 1\textwidth]{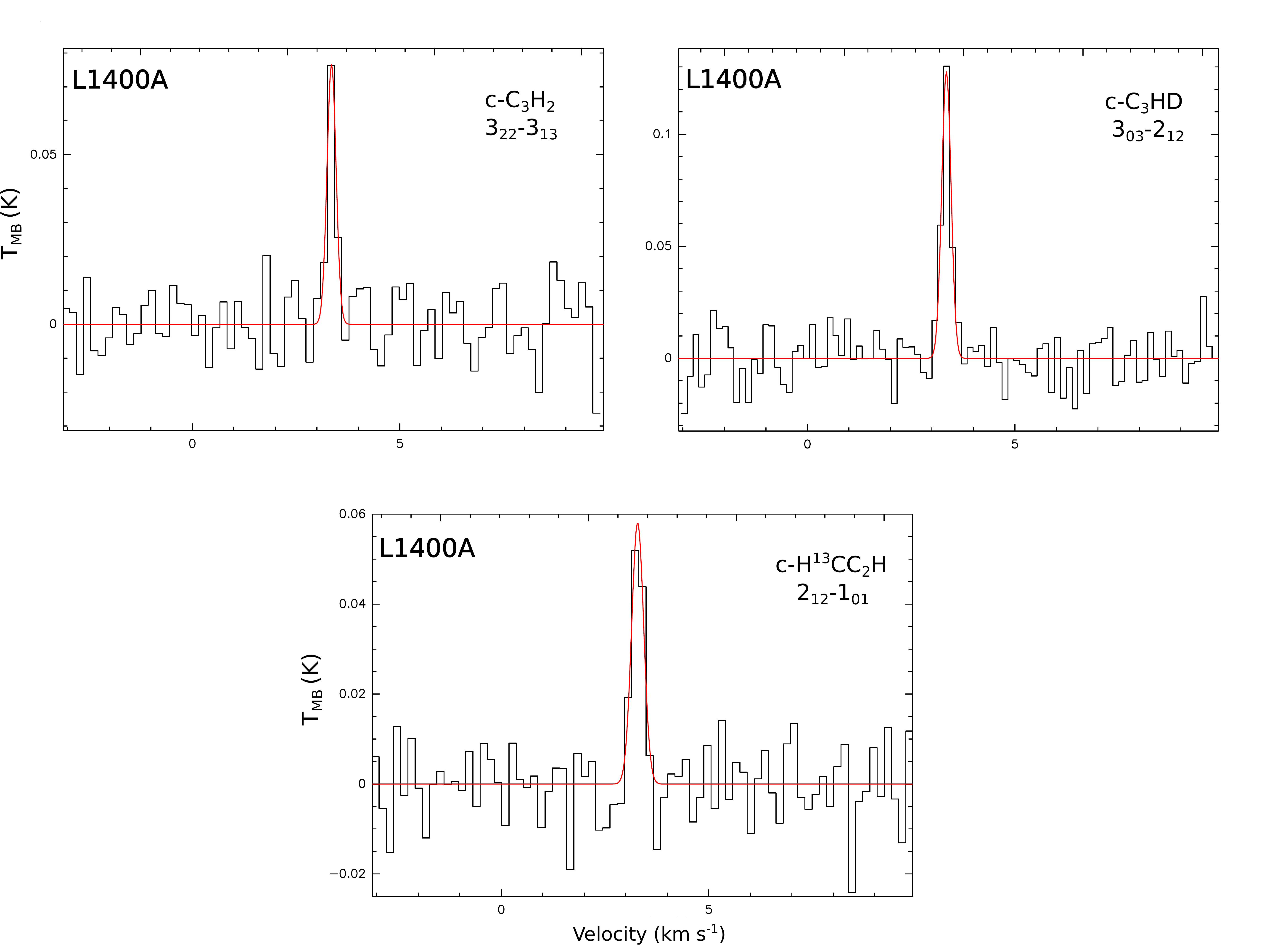}
	\caption{Spectra of the isotopologues of $c$-$\mathrm{C_{3}H_{2}}$ observed  toward the starless core L1400A. The red line plots the CLASS Gaussian fit.}
	\label{Fig:ratios7}
\end{figure*}

\begin{figure*}[h]
	\centering
	\includegraphics[width = 0.52\textwidth]{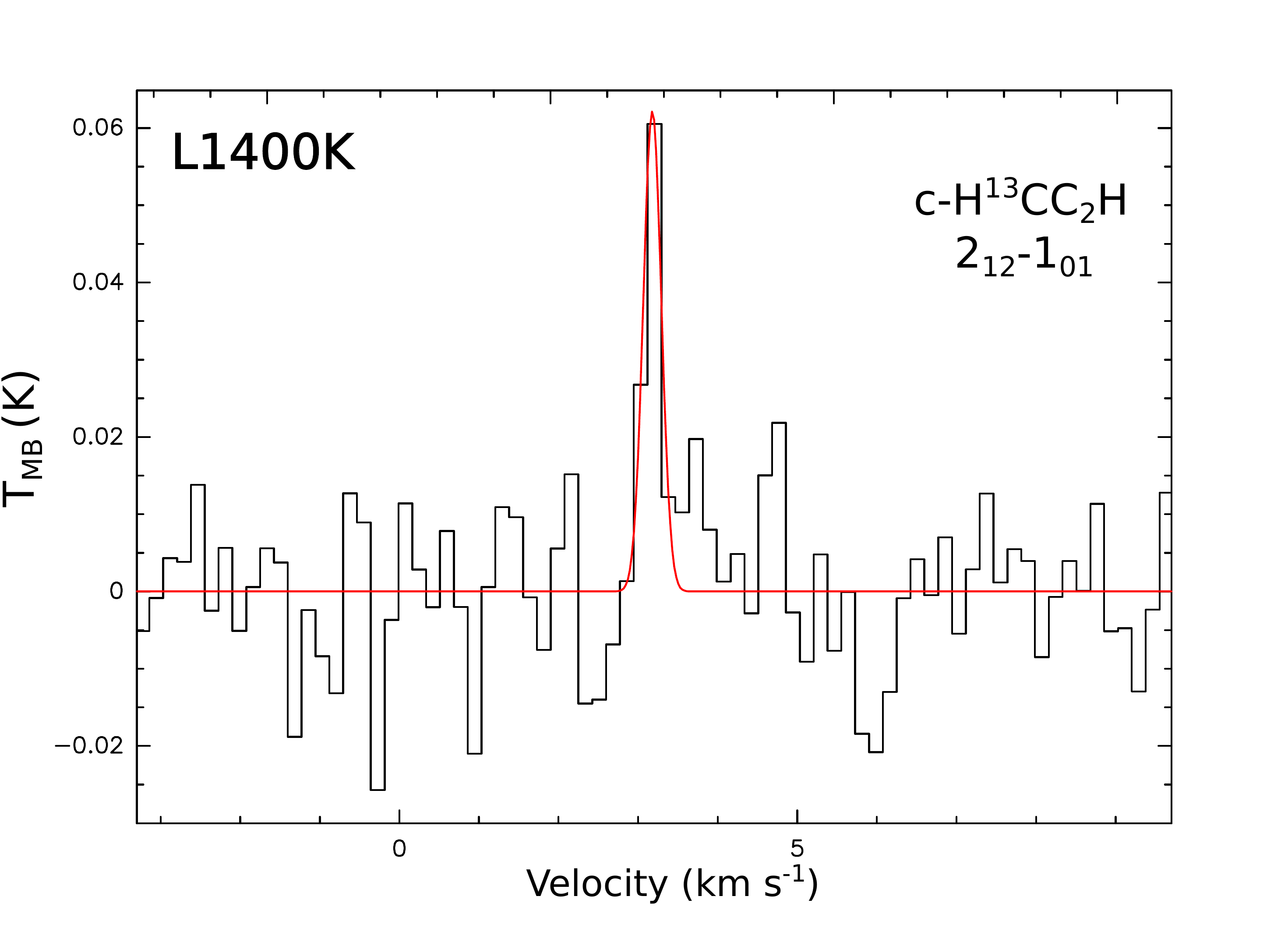}
	\caption{Spectrum of the main species $c$-$\mathrm{C_{3}H_{2}}$ observed  toward the starless core L1400K. The red line plots the CLASS Gaussian fit.}
	\label{Fig:ratios8}
\end{figure*}

\begin{figure*}[h]
	\centering
	\includegraphics[width = 1\textwidth]{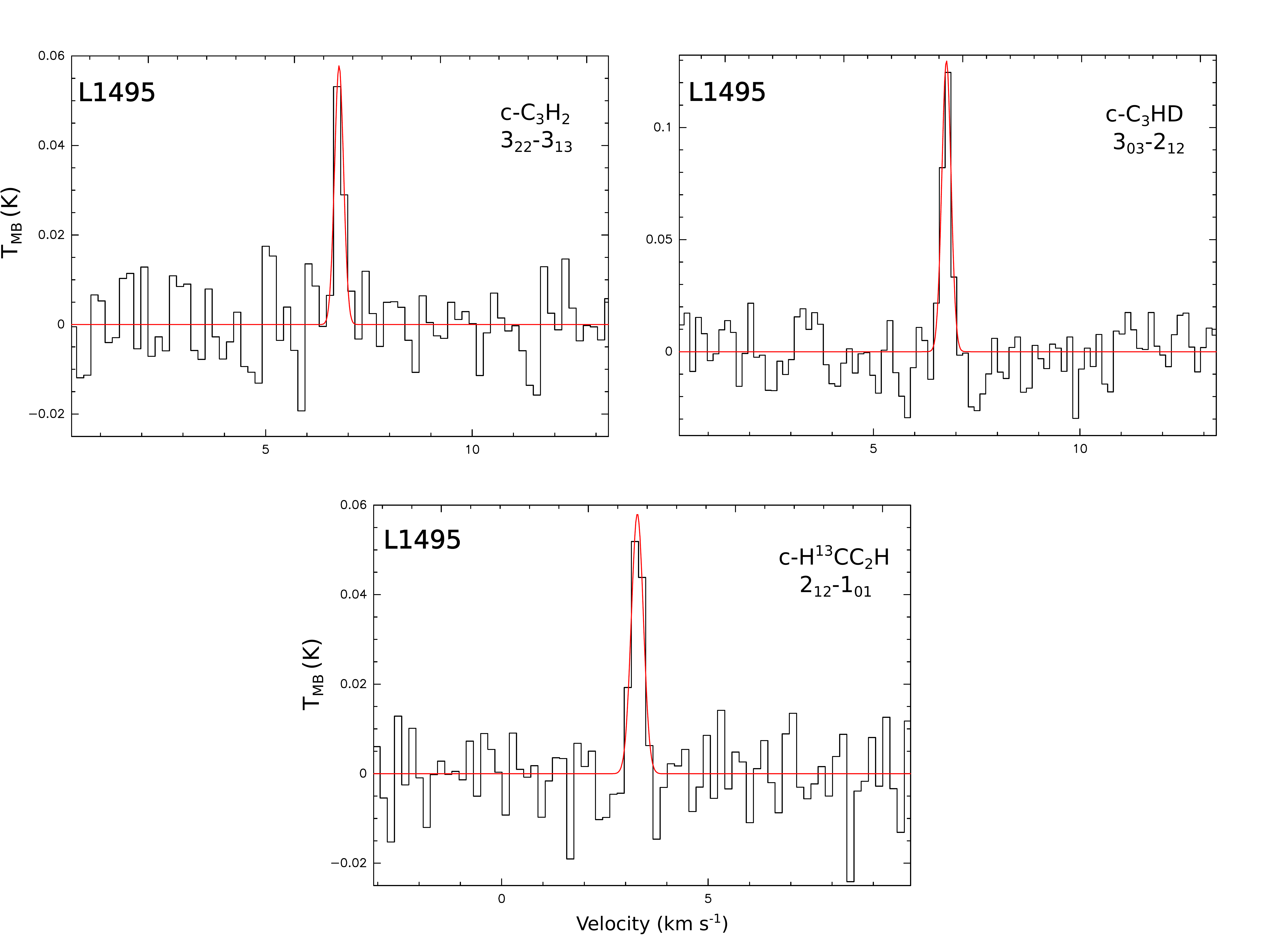}
	\caption{Spectra of the isotopologues of $c$-$\mathrm{C_{3}H_{2}}$ observed  toward the starless core L1495. The red line plots the CLASS Gaussian fit.}
	\label{Fig:ratios9}
\end{figure*}

\begin{figure*}[h]
	\centering
	\includegraphics[width = 1\textwidth]{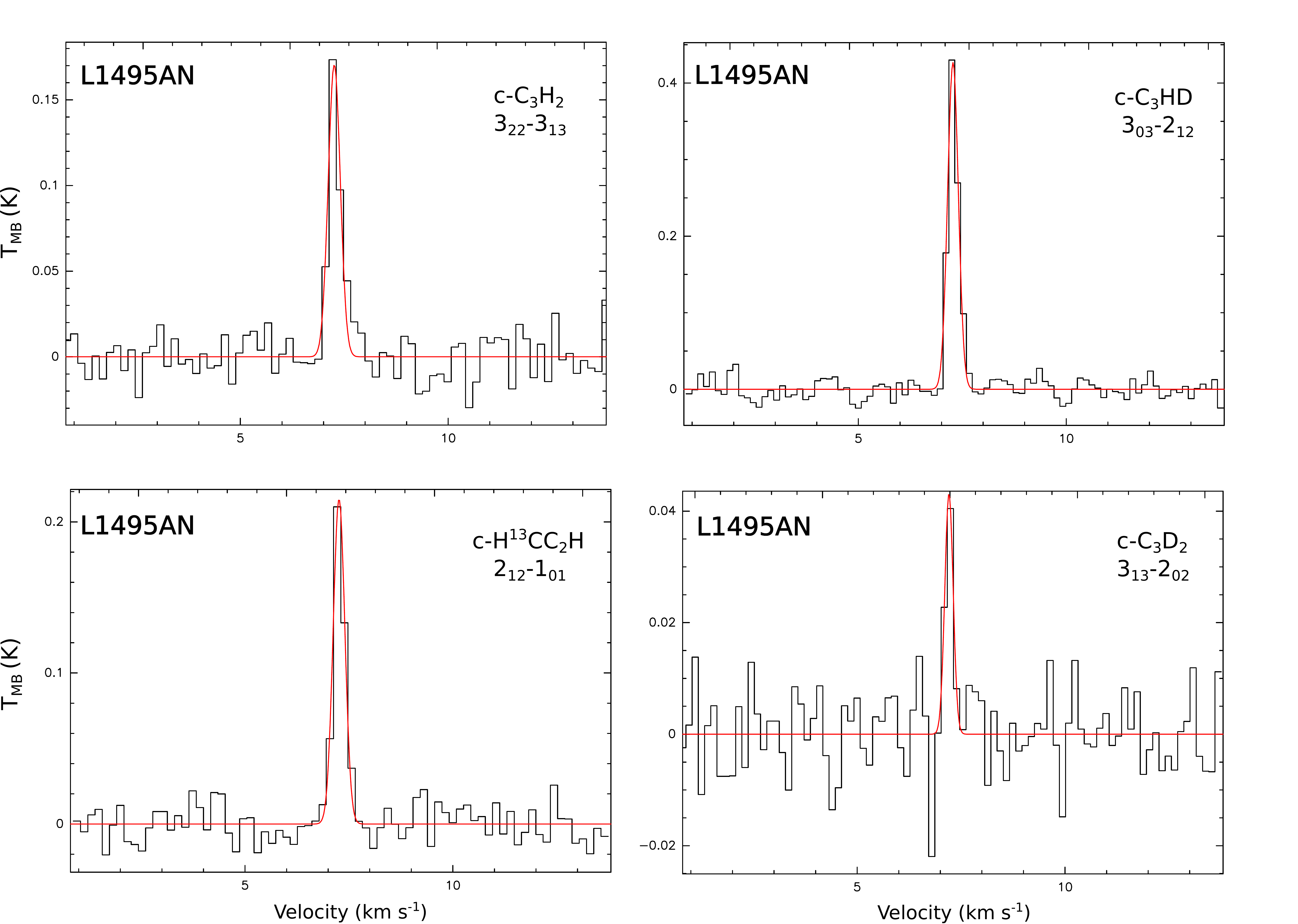}
	\caption{Spectra of the isotopologues of $c$-$\mathrm{C_{3}H_{2}}$ observed toward the starless core L1495AN. The red line plots the CLASS Gaussian fit.}
	\label{Fig:ratios10}
\end{figure*}

\begin{figure*}[h]
	\centering
	\includegraphics[width = 1\textwidth]{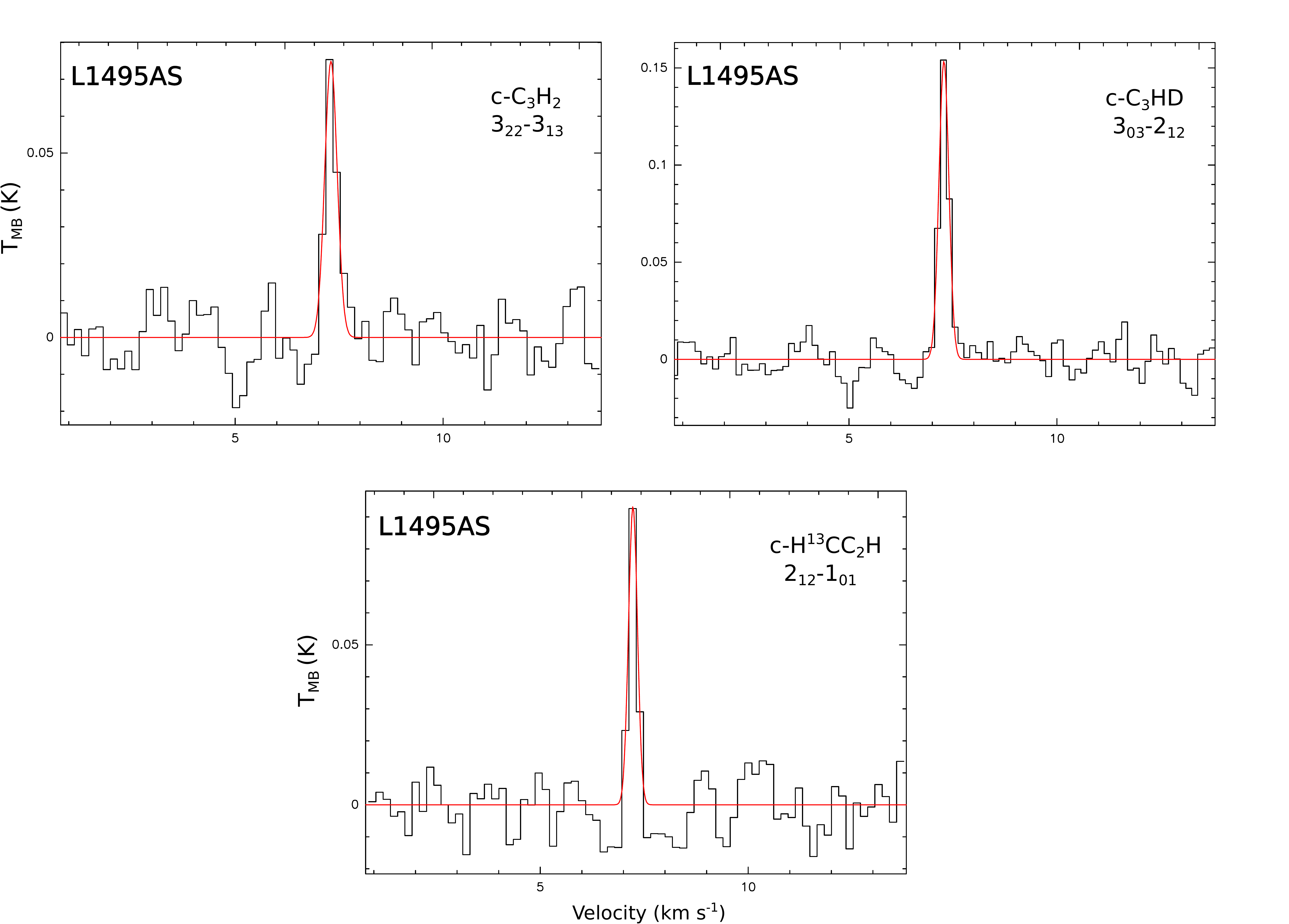}
	\caption{Spectra of the isotopologues of $c$-$\mathrm{C_{3}H_{2}}$ observed toward the starless core L1495AS. The red line plots the CLASS Gaussian fit.}
	\label{Fig:ratios11}
\end{figure*}

\begin{figure*}[h]
	\centering
	\includegraphics[width = 1\textwidth]{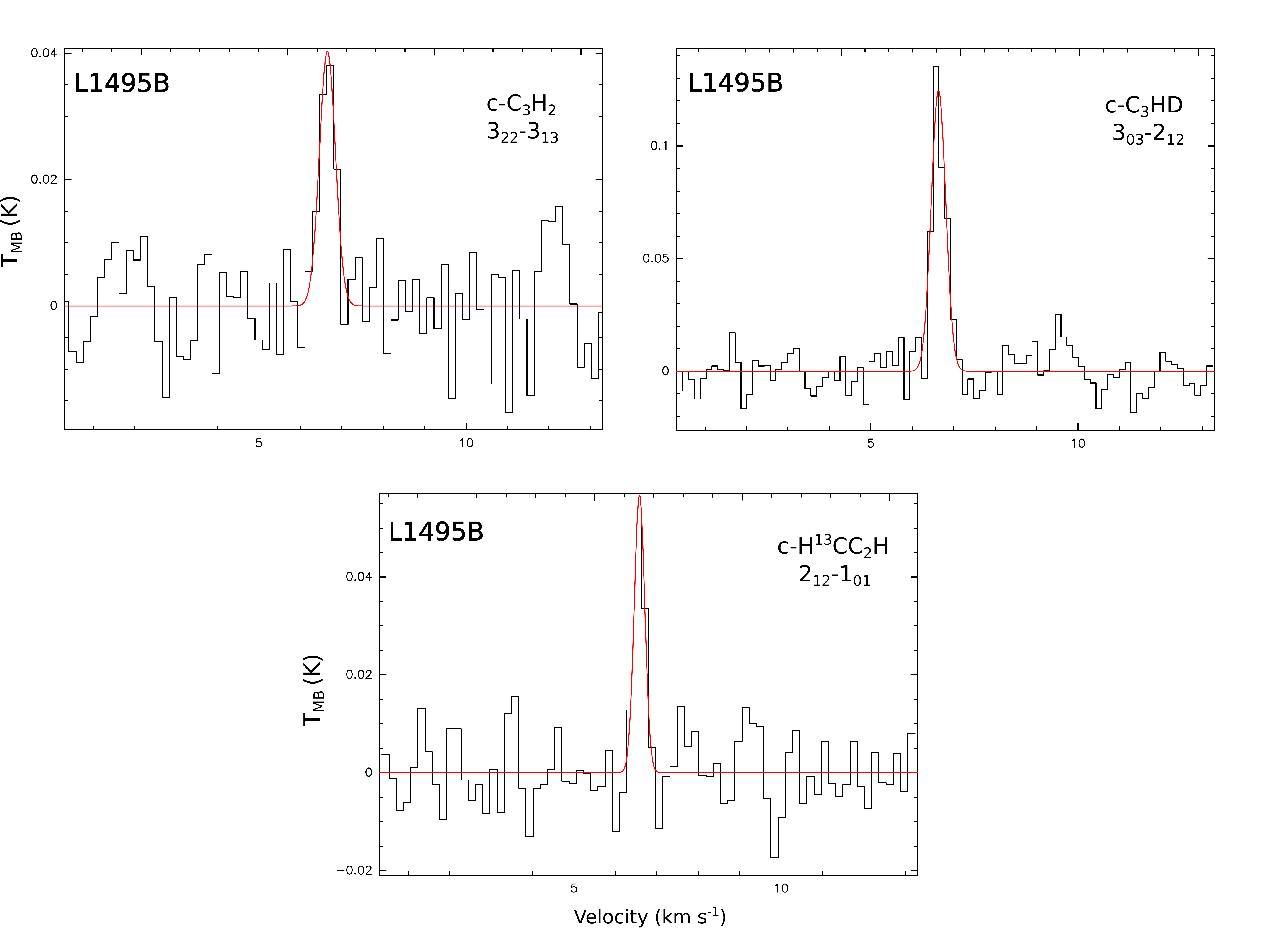}
	\caption{Spectra of the isotopologues of $c$-$\mathrm{C_{3}H_{2}}$ observed  toward the starless core L1495B. The red line plots the CLASS Gaussian fit.}
	\label{Fig:ratios12}
\end{figure*}

\begin{figure*}[h]
	\centering
	\includegraphics[width = 1\textwidth]{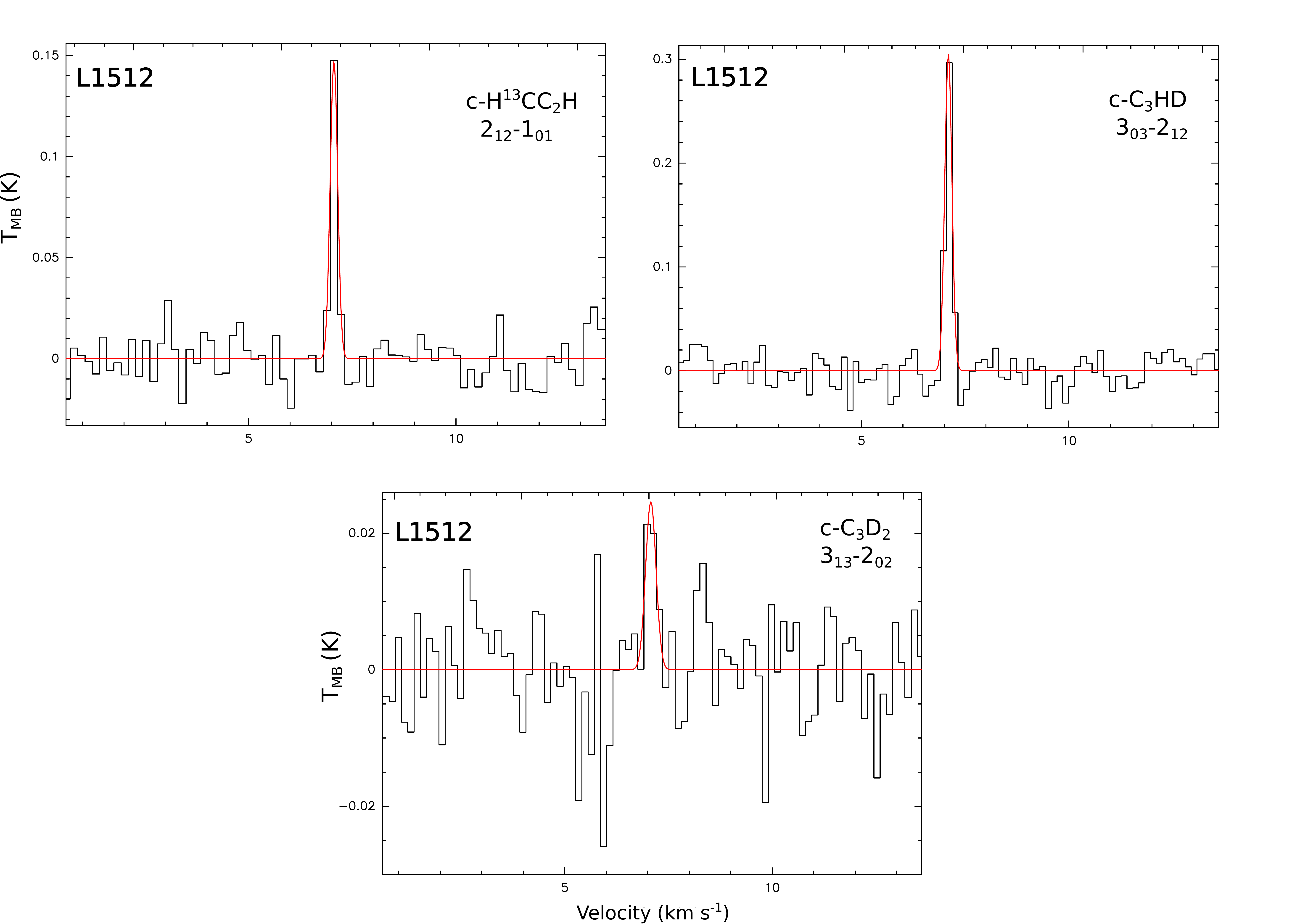}
	\caption{Spectra of the isotopologues of $c$-$\mathrm{C_{3}H_{2}}$ observed toward the starless core L1512. The red line plots the CLASS Gaussian fit.}
	\label{Fig:ratios13}
\end{figure*}

\begin{figure*}[h]
	\centering
	\includegraphics[width = 1\textwidth]{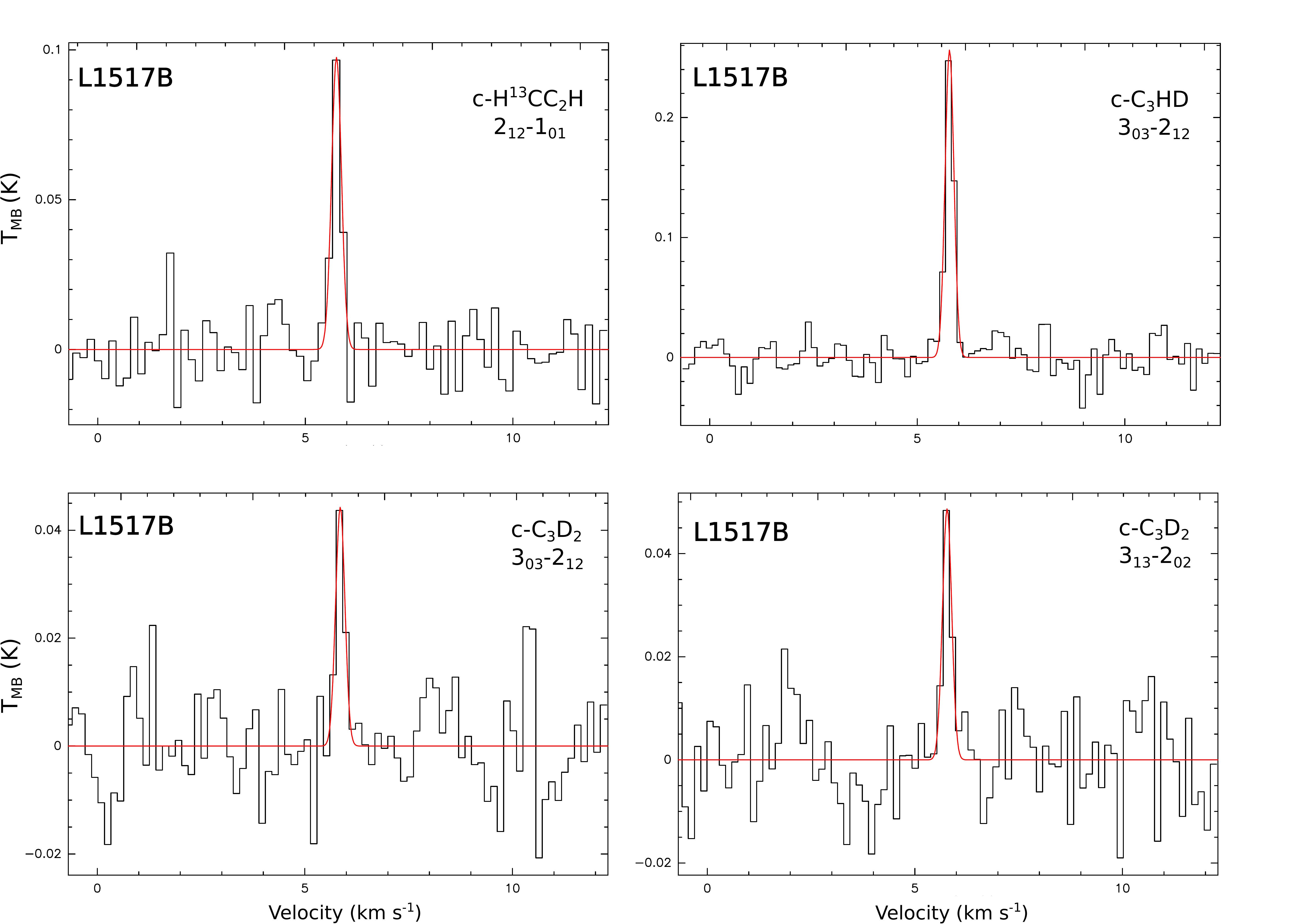}
	\caption{Spectra of the isotopologues of $c$-$\mathrm{C_{3}H_{2}}$ observed toward the starless core L1517B. The red line plots the CLASS Gaussian fit.}
	\label{Fig:ratios14}
\end{figure*}

\begin{figure*}[h]
	\centering
	\includegraphics[width = 1\textwidth]{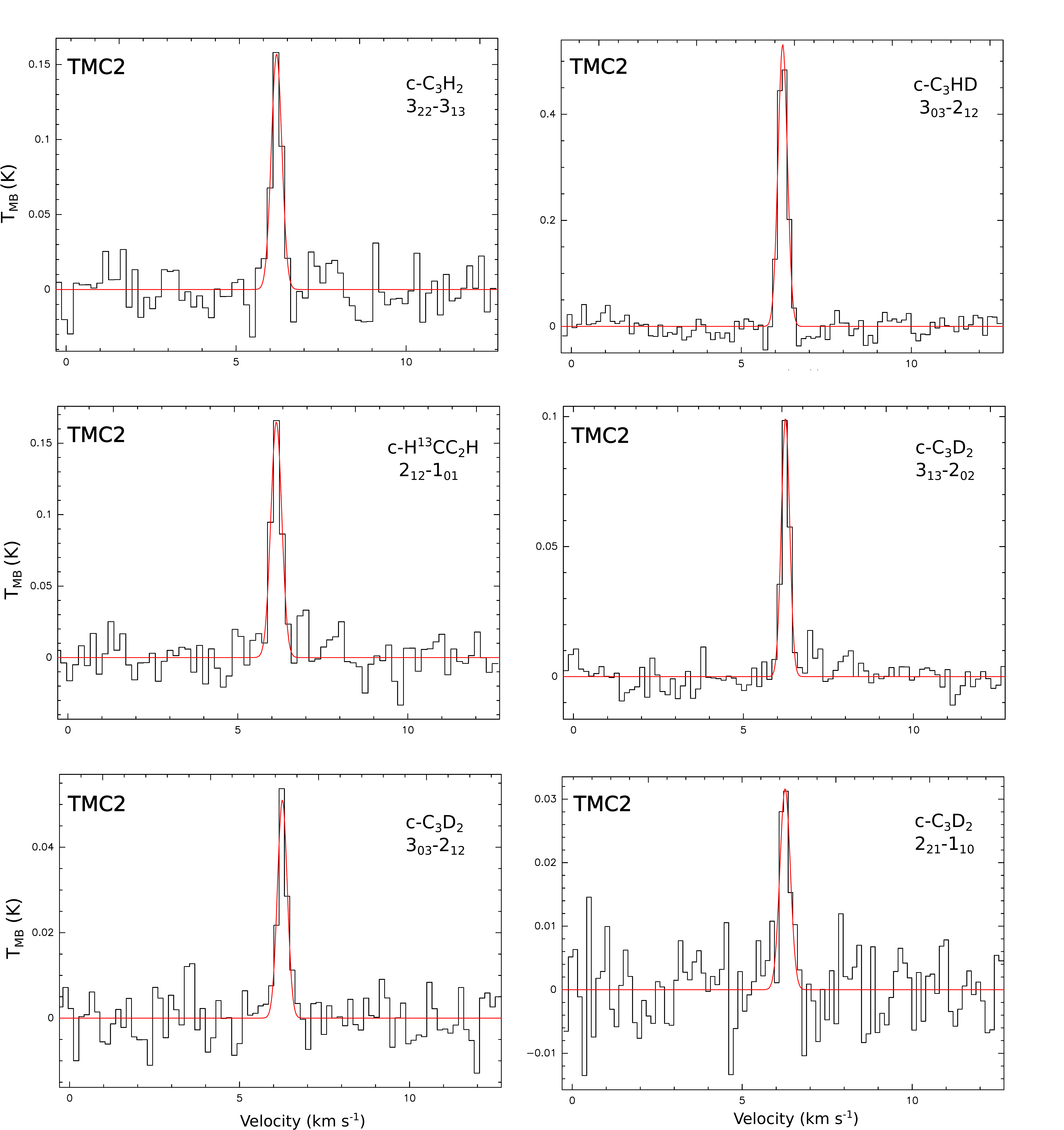}
	\caption{Spectra of the isotopologues of $c$-$\mathrm{C_{3}H_{2}}$ observed  toward the pre-stellar core TMC2. The red line plots the CLASS Gaussian fit.}
	\label{Fig:ratios16}
\end{figure*}

\clearpage

\section{The $c$-$\mathrm{C_{3}H_{2}}$ and $c$-$\mathrm{C_{3}HD}$ distribution across the pre-stellar core L1544} \label{sipila}
We use the chemical/physical model for L1544 described in \cite{sipila16} to simulate the abundances of gaseous and solid $c$-$\mathrm{C_{3}H_{2}}$, $c$-$\mathrm{C_{3}HD}$ and the ratio $c$-$\mathrm{C_{3}HD}$/$c$-$\mathrm{C_{3}H_2}$ as functions of distance from the core centre, defined by the position of the millimeter dust continuum peak. Figure \ref{Fig:model_sipila} shows the abundances of the gaseous and solid species as well as the deuteration level $c$-$\mathrm{C_{3}HD}$/$c$-$\mathrm{C_{3}H_2}$ at three different  times: $10^4$, $10^5$ and $10^6$ yr.  The depletion of $c$-$\mathrm{C_{3}H_{2}}$ and $c$-$\mathrm{C_{3}HD}$ toward the center increases with the evolution of the core, as expected. At $t=10^6$ yr the depletion zone of both species reaches a few 1000 AU. Here we confirm the fact that $c$-$\mathrm{C_{3}H_{2}}$ as well as its deuterated counterpart stop tracing the zone where high levels of deuterium fraction are present, as already suggested in \S \ref{depletion_factor}. The right panel of Figure \ref{Fig:model_sipila} shows that the total deuteration level of gaseous and solid $c$-$\mathrm{C_{3}H_{2}}$ is less than 20\% at $t=10^6$ yr. This means that one of the most advanced gas-grain chemical codes including deuterium fractionation is not able to reproduce the  large deuterium fraction of 23\% observed in $c$-$\mathrm{C_{3}H_{2}}$ toward the young protostar HH211 (which represents the next evolutionary state after the evolved pre-stellar core L1544), suggesting either that some important surface processes are missing in the current chemical scheme, or that the relative rates of the currently-included processes need to be modified. 

\begin{figure*}[h]
	\centering
	\includegraphics[width = 0.8\textwidth]{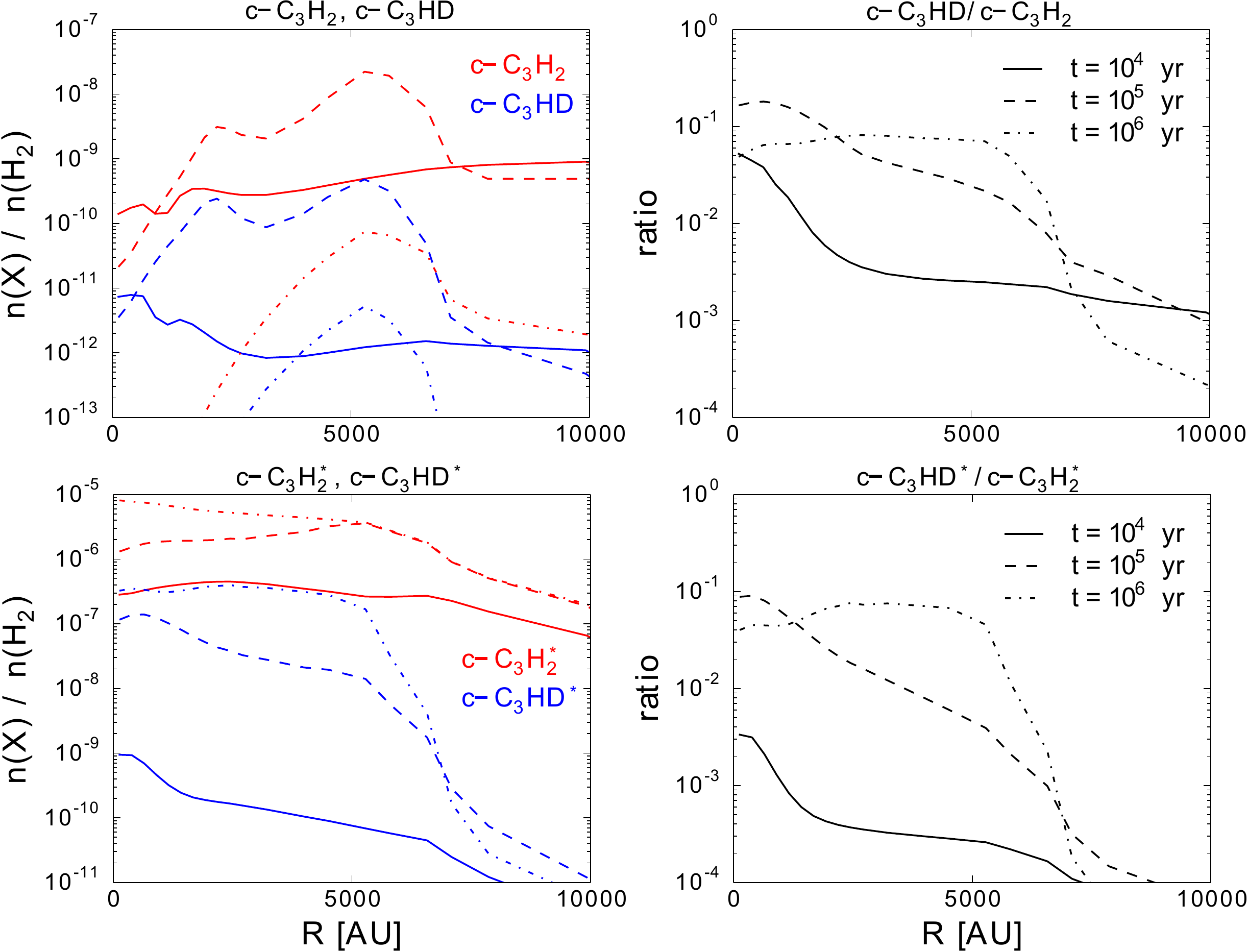}
	\caption{Abundance profiles of gaseous and solid $c$-$\mathrm{C_{3}H_{2}}$ (red), $c$-$\mathrm{C_{3}HD}$ (blue) and the ratio $c$-$\mathrm{C_{3}HD}$/$c$-$\mathrm{C_{3}H_{2}}$ toward L1544, as functions of distance, from the core center to a radius of $10^4$AU. The solid species are marked with an asterisk. The abundances and abundance ratios are plotted at three different times: $10^4$, $10^5$ and $10^6$ yr.}
	\label{Fig:model_sipila}
\end{figure*}
\clearpage

\section{Error estimation of the $\mathrm{H_2}$ column density } \label{error_est}
One source of uncertainty in the estimation of $N(\mathrm{H_2})$, which is derived from the SPIRE images at (250, 350, 500) $\mu$m, is the flux uncertainty. 
The flux calibration of the SPIRE photometer is based on Neptune. Being a bright source, Neptune produces high SNR spectra and has a well understood submillimeter spectrum. The calibration flux densities for Neptune at (250, 350, 500) $\mu$m  are (160, 100, 60) Jy and the absolute flux uncertainty  for Neptune is estimated to be 4\%, which corresponds to the absolute calibration uncertainty. Considering also the relative calibration and the extended source calibration uncertainty, the total flux uncertainty for the SPIRE bands amounts to 7\% according to the SPIRE Handbook.

The column density of  $\mathrm{H_2}$ and its total error are estimated by applying a Monte Carlo fitting procedure with 1000 iterations. After every iteration, the noise level $\epsilon_{\mathrm{noise}}$ is added to every map, pixel by pixel, following the equation:
\begin{eqnarray}
\epsilon_{\mathrm{noise}} = \epsilon_1 \cdot 0.07 \cdot I_{\nu} + \delta I_{\nu} \cdot \epsilon_2,  \nonumber
\end{eqnarray}  
where $I_{\nu}$ is the detected intensity at a frequency $\nu$, $\epsilon_{1,2}$ describes a random number taken from a standard normal distribution and  $\delta I_{\nu}$ is the statistical error on the flux density value in each pixel, produced by the pipeline.
The fitting method gives a cube of $N=1000$ maps. The fist map with $\epsilon_{\mathrm{noise}}$=0 gives the resulting  $N(\mathrm{H_2})$, while its uncertainty is given by the standard deviation of the remaining maps. 


%
\end{document}